\documentclass[acmsmall]{acmart}

\setcopyright{none}
\renewcommand\footnotetextcopyrightpermission[1]{} 

\AtBeginDocument{%
  }

\usepackage{algorithmic}
\usepackage[ruled,vlined,linesnumbered]{algorithm2e}

\usepackage{enumitem}
\usepackage{multirow}
\usepackage{diagbox}
\usepackage{tcolorbox}
\usepackage{xurl}

\makeatletter
\DeclareRobustCommand\onedot{\futurelet\@let@token\@onedot}
\def\@onedot{\ifx\@let@token.\else.\null\fi\xspace}

\def\eg{\emph{e.g}\onedot} 
\def\ie{\emph{i.e}\onedot}

\def\etal{\emph{et al}\onedot}
\makeatother

\usepackage{xspace}
\def \projectName {GradSplitter\xspace}

\usepackage{colortbl}

\begin{document}

\title{Reusing Convolutional Neural Network Models through Modularization and Composition}

\author{Binhang Qi}
\affiliation{%
  \institution{SKLSDE, School of Computer Science and Engineering, Beihang University}
  \country{China}
}
\email{binhangqi@buaa.edu.cn}

\author{Hailong Sun}
\authornote{Corresponding author. Hailong Sun is also with Hangzhou Innovation Institute, Beihang University, China}
\affiliation{
  \institution{SKLSDE, School of Software, Beihang University}
  \country{China}
}
\email{sunhl@buaa.edu.cn}

\author{Hongyu Zhang}
\affiliation{%
 \institution{Chongqing University}
 \country{China}
}
\email{hyzhang@cqu.edu.cn}

\author{Xiang Gao}
\affiliation{%
  \institution{School of Software, Beihang University}
  \country{China}
}
\email{xiang\_gao@buaa.edu.cn}

\begin{abstract}
With the widespread success of deep learning technologies, many trained deep neural network (DNN) models are now publicly available. 
However, directly reusing the %
public DNN models for new tasks often fails due to mismatching functionality or performance. 
Inspired by the notion of modularization and composition in software reuse, we investigate the possibility of improving %
the reusability of DNN models in a more fine-grained manner. 
Specifically, we propose two modularization approaches named CNNSplitter and \projectName, which can decompose a trained convolutional neural network (CNN) model for $N$-class classification into $N$ 
small reusable modules. 
Each module recognizes one of the $N$ classes and contains a part of the convolution kernels of the trained CNN model. 
Then, the resulting modules can be reused to patch existing CNN models or build new CNN models through composition.
The main difference between CNNSplitter and GradSplitter lies in their search methods: the former relies on a genetic algorithm to explore search space, while the latter utilizes a gradient-based search method.
Our experiments with three representative CNNs on three widely-used public datasets demonstrate the effectiveness of the proposed approaches. %
Compared with CNNSplitter, GradSplitter incurs less accuracy loss, produces much smaller modules (19.88\% fewer kernels), and achieves better results on patching weak models. %
In particular, experiments on GradSplitter show that (1) by patching weak models, the average improvement in terms of precision, recall, and F1-score is 17.13\%, 4.95\%, and 11.47\%, respectively, and
(2) for a new task, compared with the models trained from scratch, reusing modules achieves similar accuracy (the average loss of accuracy is only 2.46\%) without a costly training process. 
Our approaches provide a viable solution to the rapid development and improvement of CNN models.
\end{abstract}


\maketitle

\thispagestyle{plain}
\pagestyle{plain}

\section{Introduction}
\label{sec:intro}

Modularization and composition are fundamental concepts in software engineering,
which facilitate software development, reuse, and maintenance by dividing an entire software system into a set of smaller modules.
Each module is capable of carrying out a certain task and can be composed with other modules~\cite{david_1,david_2, david_3}.
For instance, when debugging a buggy program, testing and patching the module that contains the bug will be much easier than analysing the entire program.

We highlight that modularization and composition are also important for deep neural networks (DNNs), such as the convolutional neural network (CNN) that is one of the most effective DNNs
for processing a variety of tasks~\cite{alexnet,girshick2014rich,long2015fully}.
Due to the widespread application of CNNs, many trained CNN models are now publicly available, and reusing existing trained models has gained increasing attention recently~\cite{fse2020modularity,nnmodularity2022icse,cnnsplitter}.
However, reusing existing models has two main challenges: (1) existing CNN models from old projects may perform unsatisfactorily in the target task, and (2)  models that can solve the target tasks may not exist.
To improve the accuracy of weak CNN models, developers often retrain the models using new data, model structures, training strategies, or hyperparameter values. Additionally, to obtain a new CNN model for a new project, developers can evaluate the accuracy of public CNN models on their own test data and choose the model with the highest accuracy for reuse.
However, current practice in model development and improvement has the following limitations: 
(1) as the neural networks are getting deeper and the numbers of parameters and convolution operations are getting larger, the time and computational cost required for training the CNN models are rapidly growing.
(2) even if existing models satisfy developers' requirements, directly reusing the model with the highest overall accuracy may not always be the best solution. For instance, the model with the highest overall accuracy may be less accurate in recognizing a certain class than other models.

At a conceptual level, a CNN model is analogous to a program~\cite{ma2018mode, pei2017deepxplore,xie2019deephunter}. 
Inspired by the application of modularization and composition in software development and debugging, it is natural to ask: \textit{can the concepts of modularization and composition be applied to CNN models for facilitating the development and improvement of CNN models?} 
Through modularization and composition, the weak modules in a weak CNN model can be identified and patched separately; thus, the weak model can be improved without costly retraining the entire model. 
Moreover, some modules can be reused to create a new CNN model without costly retraining. 
Also, a module is much smaller (\ie has fewer weights) than the entire model, which is essential for reducing the overhead of model reuse.

However, decomposing a CNN model into modules faces two main challenges:
(1) CNN models are constructed with uninterpretable weight matrices, unlike software programs, which are composed of readable statements.
Decomposing CNN models into distinct modules is challenging without fully comprehending the effect of each weight.
(2) identifying the relations between neurons and prediction tasks is difficult as the connections between neurons in a CNN are complex and dense. 
To this end, Pan \etal~\cite{fse2020modularity} proposed decomposing a fully connected neural network (FCNN) model for $N$-class classification into $N$ modules, one for each class in the original model. 
They achieved model decomposition through \textit{uncompressed modularization}, which removes individual weights from a trained FCNN model and results in modules with sparse weight matrices (to be discussed in Section \ref{subsec:discuss}). 
However, this approach~\cite{fse2020modularity} cannot be applied to other advanced DNN models like CNN models due to the weight sharing~\cite{cnn2018overview,lenet} in CNNs.
That is, different from FCNNs where the relationship between weights and neurons is many-to-one, the relationship in CNNs is many-to-many.
Removing weights for one neuron in CNNs will also affect all other neurons.
Although the follow-up work~\cite{nnmodularity2022icse} can be applied to decompose CNN models, it is still an uncompressed modularization approach.
In uncompressed modularization, a module with a sparse weight matrix has the same size as the trained model, resulting in a significant overhead of module reuse.

To address the above challenges, we propose the first \textit{compressed modularization} approach called $CNNSplitter$, which applies a genetic algorithm to decompose a CNN model into smaller and separate modules.
By {compressed modularization}, we mean removing convolution kernels
instead of individual weights, resulting in modules with smaller weight matrices. 
Inspired by the finding that different convolution kernels learn to extract different features from the data~\cite{cnn2018overview}, we generate modules by selecting different convolution kernels in a CNN model. Therefore, different from the existing work~\cite{fse2020modularity}, CNNSplitter decomposes a trained $N$-class CNN model (\textit{TM} for short) into $N$ CNN modules by removing unwanted convolution kernels. 
To decompose a trained CNN model for $N$-class classification, we formulate the modularization problem as a search problem. 
Search-based algorithms have been proven to be very successful in solving software engineering problems~\cite{harman2001search,li2016value}.
Given a space of candidate solutions, search-based approaches usually search for the optimal solution over the search space according to a user-defined objective function.
In the context of model decomposition, the candidate solution is defined as a set of sub-models containing a part of the CNN model's kernels, while the search objective is to search $N$ sub-models (as $N$ modules) with each of them recognizing one class.
To search for optimal modules, CNNSplitter employs a genetic algorithm that utilizes a combination of modules' accuracy and the difference between modules as the objective function.
In this way, a trained CNN model is decomposed into $N$ modules. %
Each module corresponds to one of the $N$ classes and %
evaluates whether or not an input belongs to the corresponding class.

\begin{figure}
	\includegraphics[width=\columnwidth]{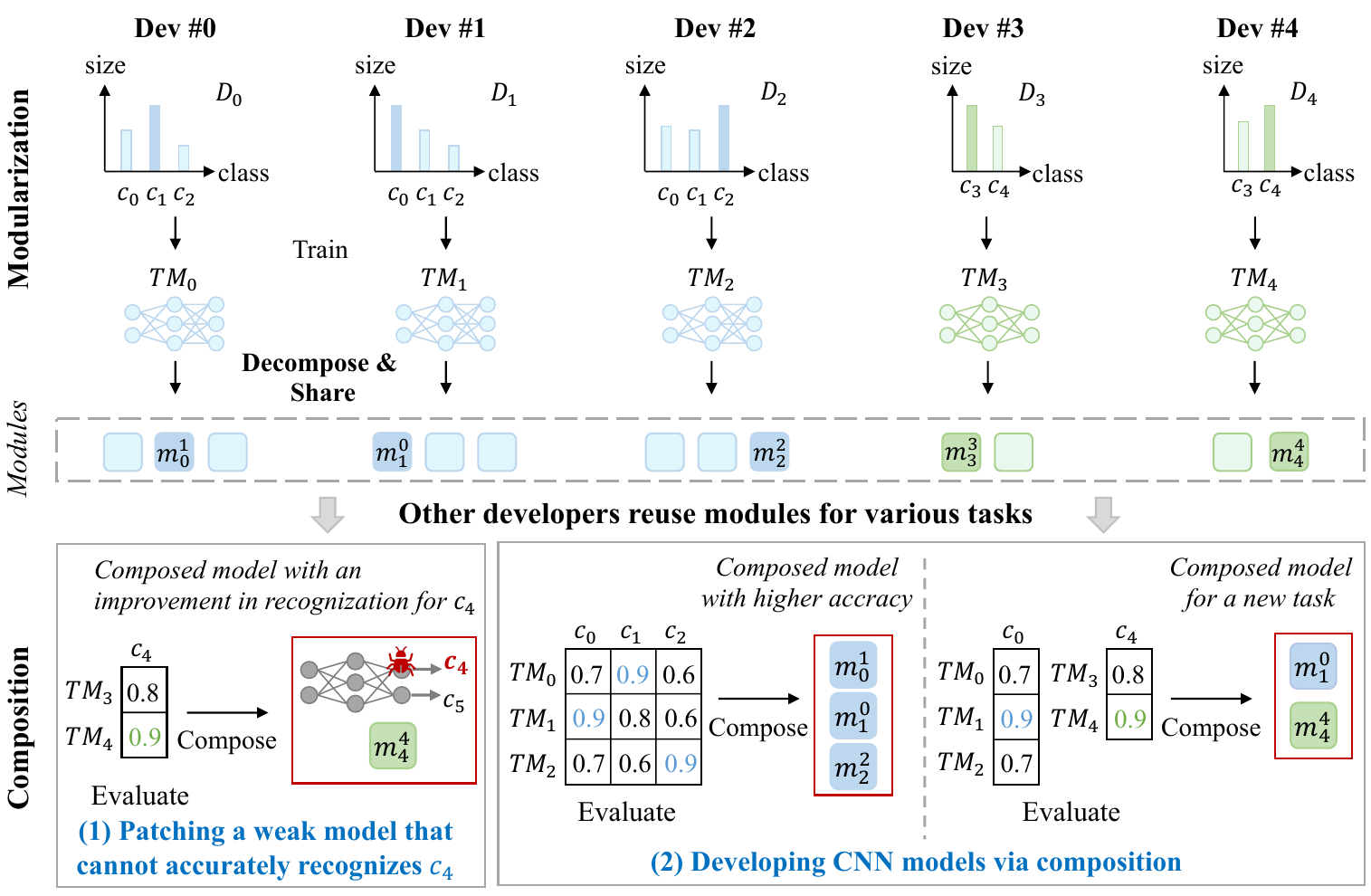}
    \caption{An illustration of model modularization and composition.}%
    \label{fig:motivation}
    \vspace{-6pt}
\end{figure}

Due to the huge search space, the traditional search approach (\eg genetic algorithm) could incur a large time cost. 
To improve the efficiency of modularization, we further propose a gradient-based compressed modularization approach named $GradSplitter$, which applies a gradient-based search method to explore the search space. 
To decompose a \textit{TM} into $N$ modules, GradSplitter initializes a \textit{mask} and a \textit{head} for each module.
The \textit{mask} consists of an array of 0s and 1s, where 0 (or 1) indicates that the corresponding convolution kernels in the \textit{TM} are removed (or retained). 
The \textit{head}, consisting of fully connected (FC) layers, is appended after the output layer of the masked \textit{TM} to convert $N$-classification to binary classification, \ie whether an input belongs to the class of the corresponding module.
Then, a module $M$ is created by $M = mask \odot TM \oplus head$, where $\odot$ denotes the removal of the corresponding kernels from \textit{TM} according to the 0s in the $mask$ and $\oplus$ denotes the appending of the \textit{head} after the masked \textit{TM}. 
\projectName combines the outputs of $N$ modules and optimizes the masks and heads of $N$ modules jointly on the $N$-class classification task.
For the optimization, a gradient descent approach is used to minimize the number of retained kernels and the cross-entropy between the predicted class and the actual class.
In this way, GradSplitter can search for $N$
modules with each of them containing only relevant kernels.

As illustrated in Figure \ref{fig:motivation}, third-party developers, labeled as ``Dev \#0-4'', generally train models for various tasks, such as distinct tasks (e.g., trained models $TM_2$ and $TM_3$) or similar tasks but using datasets with different distributions (e.g., $TM_0$, $TM_1$, and $TM_2$). With the modularization technique, developers not only release their trained CNN models ($TMs$ for short) but also share a set of smaller and reusable modules decomposed from $TMs$. With the shared modules, similar to reusing complete models, other developers can evaluate and reuse the suitable modules according to their demands without costly training. For instance, to improve the recognition of a weak CNN model on a target class, the module with the best performance (\eg F1 score) in classifying the target class is reused as a patch to be combined with the weak CNN model. Additionally, developers can build new CNN models entirely by combining optimal modules. Consequently, composed models (\textit{CMs} for short) are constructed through module reuse, which can address new tasks or achieve better performance than existing trained models.

We evaluate CNNSplitter and \projectName using three representative CNNs with different structures %
on three widely-used datasets (CIFAR-10~\cite{cifar10}, CIFAR-100~\cite{cifar10}, and SVHN~\cite{svhn}). 
The experimental results show that by decomposing a \textit{TM} into modules with \projectName 
and then combining the modules to build a \textit{CM} that is functionally equivalent to the \textit{TM}, only a negligible loss of accuracy (0.58\% on average) is incurred.
In addition, each module retains only 36.88\% of the convolution kernel of the \textit{TM} on average. 
To validate the effectiveness of module reuse, we apply modules as patches to improve three common types of weak CNN models, \ie overly simple model, underfitting model, and overfitting model. 
Overall, after patching, the averaged improvements in terms of precision, recall, and F1-score are 17.13\%, 4.95\%, and 11.47\%, respectively.
Also, for a new task, we develop a \textit{CM} entirely by reusing the modules with the best performance in the corresponding class. Compared with the models retrained from scratch, the \textit{CM} achieves similar accuracy with a loss of only 2.46\%. Even though there may exist \textit{TMs} that can be directly reused, the \textit{CM} outperforms the best \textit{TM} and the average improvement in accuracy is 5.18\%.
Although modularization and composition incur additional time and GPU memory overhead, the experimental results demonstrate that the overhead is affordable. 
In particular, \textit{CMs} can make prediction faster than \textit{TMs} by executing modules in parallel and incur 28.6\% less time overhead than \textit{TMs}.

The main contributions of this work are as follows:
\begin{itemize}[leftmargin=*]
    \item We propose compressed modularization approaches including CNNSplitter and \projectName, which can decompose a CNN model into a set of reusable modules.
    We also apply CNNSplitter and \projectName to improve CNN models and build CNN models for new tasks through module reuse.
    To our best knowledge, CNNSplitter is the first \textit{compressed modularization} approach that 
    can decompose trained CNN models into CNN modules and reduce the overhead of module reuse.
    \item We formulate the modularization of CNNs as a search problem and design a genetic algorithm and a gradient descent-based search method to solve it. %
    Especially, we propose three heuristic methods to alleviate the problem of excessive search space and time complexity in CNNSplitter and design a module evaluation method to recommend the optimal module for module reuse.
    \item We conduct extensive experiments using three representative CNNs on three widely-used datasets. The results show that CNNSplitter and \projectName can decompose a trained CNN model into modules with negligible loss of model accuracy. Also, 
    the experiments demonstrate the effectiveness of developing accurate CNN models by reusing modules. 
\end{itemize}

This work is an extension of our early work published as a conference paper~\cite{qi2022patching}, %
in which we proposed a compressed modularization approach, CNNSplitter, by applying a genetic algorithm to decompose CNN models into smaller CNN modules. The experiments demonstrated that CNNSplitter could be applied to patch weak CNN models through reusing modules obtained from strong CNN models, thus improving the recognition ability of the weak CNN models. 
Compared to~\cite{qi2022patching}, this work (1) proposes a novel compressed modularization approach, named GradSplitter, which applies a gradient-based search method to decompose CNN models and outperforms CNNSplitter in both efficiency and effectiveness (see Sec. \ref{sec:approach_decom}), (2) verifies the effectiveness of CNN modularization and composition in a new application (see Sec. \ref{sec:approach_reuse}), and (3) conducts more comprehensive experiments to evaluate the effectiveness and efficiency of CNN modularization and composition (see RQ3 to RQ5 in Sec. \ref{subsec:exp_result}).

\textbf{Replication Package:} Our source code and experimental data are available at \cite{cnnsplitter} and \cite{gradsplitter}.

\label{subsec:cnn}
\begin{figure}[t]
	\centering
	\includegraphics[width=8.3cm]{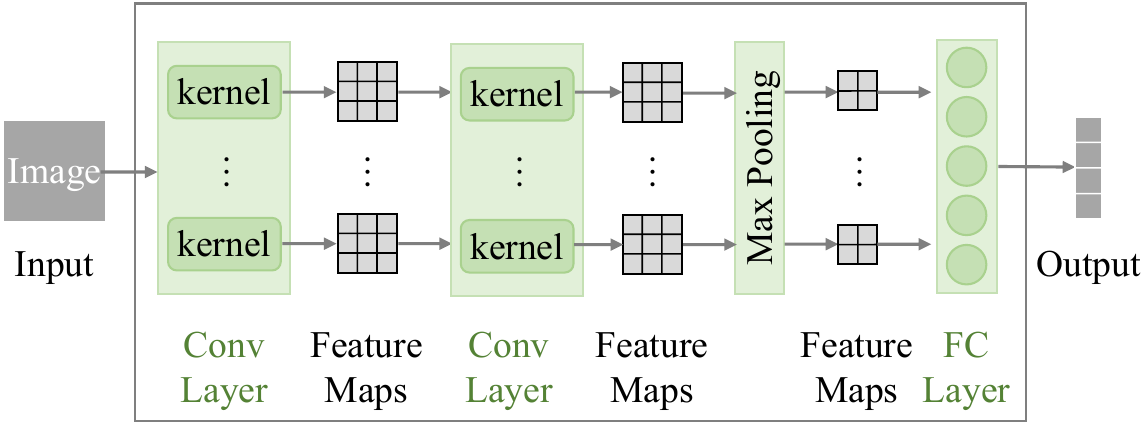}
    \caption{The architecture of a typical CNN model.}
    \label{fig:cnn-model}
    \vspace{-6pt}
\end{figure}

\section{Background}
\label{sec:background}
This section briefly introduces some preliminary information about this study, including the convolutional neural network and genetic algorithm.

\subsection{Convolutional Neural Network}
A CNN model typically contains convolutional layers, pooling layers, and FC layers, of which the convolutional layers are the core of a CNN~\cite{cnn2018overview,vgg}.
For instance, Figure \ref{fig:cnn-model} shows the architecture of a typical CNN model. 
A convolutional layer contains many convolution kernels, each of which learns to extract a local feature of an input tensor~\cite{cnn2018overview,lenet}. 
An input tensor can be an input image or a feature map produced by the previous convolutional layer or pooling layer. 
A pooling layer provides a downsampling operation~\cite{cnn2018overview}. For instance, max pooling is the most popular form of pooling operation, which reduces the dimensionality of the feature maps by extracting the maximum value and discarding all the other values.
FC layers are usually at the end of CNNs and are used to make predictions based on the features extracted from the convolutional and pooling layers.

\begin{figure}
    \centering
    \includegraphics[width=7.9cm]{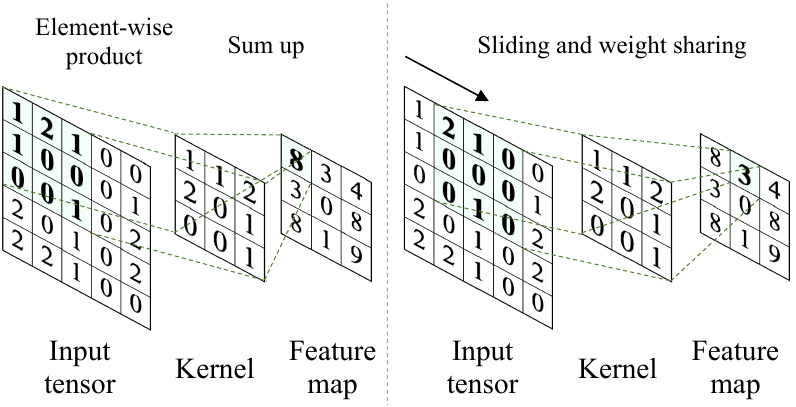}
    \caption{An example of convolution operation.}
    \label{fig:conv_oper}
    \vspace{-6pt}
\end{figure}

Figure \ref{fig:conv_oper} shows an example of convolution operation. By sliding over the input tensor, a convolution kernel calculates 
how well the local features on the input tensor match the feature that the convolution kernel learns to extract. 
The more similar the local features are to the features extracted by the convolution kernel, the larger the output value of the convolution kernel at the corresponding position, and vice versa. 
Since a convolution kernel slides over the input tensor to match features and produce a feature map, all values in the feature map share the same convolution kernel.
For instance, in the feature map of Figure~\ref{fig:conv_oper}, the values in the top-left (8) and top-middle (3) share the same kernel, \ie weights.
Weight sharing~\cite{cnn2018overview,lenet} is one of the key features of a convolutional layer. 
The values in a feature map reflect the degree of matching between the kernel and the input tensor. For instance, compared to the position in the input tensor corresponding to the top-middle (3) in the feature map, the position corresponding to the top-left (8) is more similar to the kernel.

\subsection{Genetic Algorithm}
\label{subsec:ga}
Inspired by the natural selection process, the genetic algorithm performs \textit{selection}, \textit{crossover}, and \textit{mutation} for several \textit{generations} (\ie rounds) to generate solutions for a search problem~\cite{houck1995genetic, reeves1995genetic}.
A standard genetic algorithm has two prerequisites, \ie the representation of an \textit{individual} and the calculation of an individual's \textit{fitness}.
For instance, the genetic algorithm is used to search for high-quality 
CNN architectures~\cite{evolution2017genetic, genetic2019}.
An \textit{individual} is a bit vector representing a NN architecture~\cite{evolution2017genetic}, where each bit corresponds to a convolution layer. 
The \textit{fitness} of an individual is the classification accuracy of a trained CNN model with the architecture represented by the individual. 
During the search, in each generation, the \textit{selection} operator compares the fitness of individuals and preserves the strong ones as parents that obtain high accuracy. 
The \textit{crossover} operator swaps part of two parents.
The \textit{mutation} operator randomly changes several bit values in the parents to enable or disable these convolution layers corresponding to the changed bit values. 
After the three operations, a new \textit{population} (\ie a set of individuals) is generated. And the process continues with the new generation iteratively until it reaches a fixed number of generations or an individual with the target accuracy is obtained.

\section{CNNSplitter: Genetic algorithm-based Compressed Modularization}
\label{sec:approach_cnnsplitter}
\begin{figure*}[t]
	\centering
	\includegraphics[width=\columnwidth]{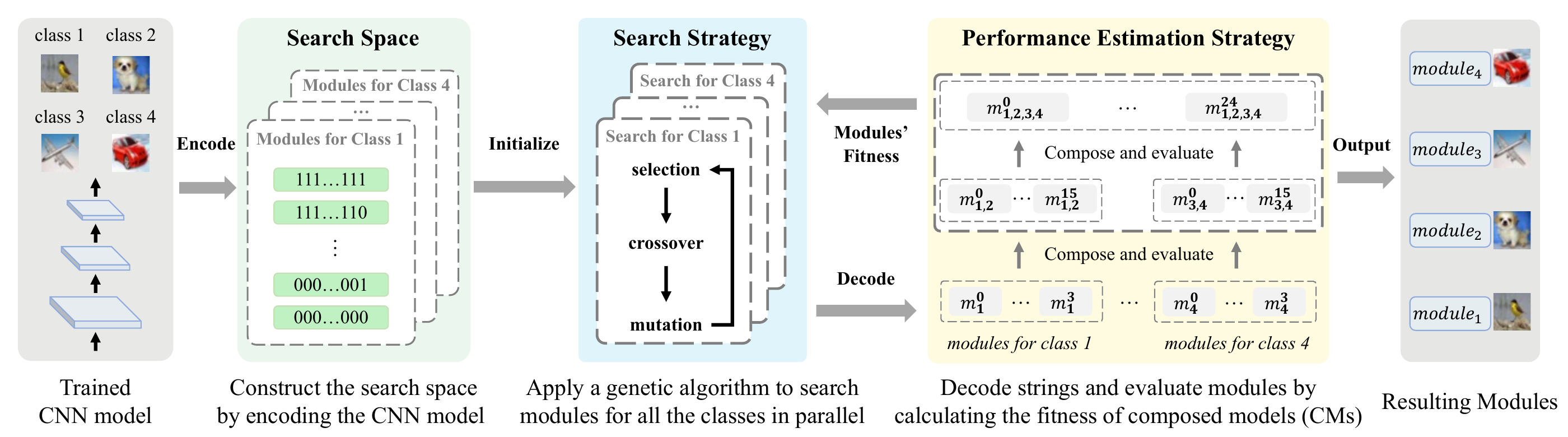}
	\vspace{-6pt}
    \caption{The overall workflow of CNNSplitter.}
    \label{fig:framework}
    \vspace{-6pt}
\end{figure*}

Figure \ref{fig:framework} shows the overall workflow of CNNSplitter. For a given trained $N$-class CNN model $\mathcal{M}{=}\{k_0, k_1,\dots, k_{L-1}\}$ with $L$ convolution kernels, the modularization process is summarized as follows: 

(1) \textit{Construction of Search Space}: CNNSplitter encodes each candidate module into a fixed-length bit vector, 
where each bit represents whether the corresponding kernels are kept or not.
The bit vectors of all candidate modules constitute the search space.

(2) \textit{Search Strategy}: From the search space, the search strategy employs a genetic algorithm to find modules for $N$ classes.

(3) \textit{Performance Estimation}: The performance estimation strategy measures the performance (\ie fitness) of the searched candidate and guides the search process. %

\subsection{Search Space}

As shown in Figure \ref{fig:framework}, the search space is represented using a set of bit vectors. %
For a CNN model with a lot of kernels, the size of vector could be very long, resulting in an excessively large search space, which could seriously impair the search efficiency. 
For instance, 10-class VGGNet-16~\cite{vgg} includes 4,224 kernels, so the number of candidate modules for each class is $2^{4224}$. 
In total, the size of the search space will be $10 \times 2^{4224}$. 
To reduce the search space, the kernels in a convolutional layer are divided into groups.
A simple way is to randomly group kernels; however,  
this could result in a group containing both kernels necessary for a module to recognize a specific class and those that are unnecessary. 
The randomness introduced by random grouping cannot be eliminated by subsequent searches, resulting in unnecessary kernels in the searched modules.

To avoid unnecessary kernels as much as possible, an \textit{importance-based grouping scheme} is proposed to group kernels based on their importance. 
As introduced in Section \ref{sec:background}, the values in a feature map can reflect the degree of matching between a convolution kernel and an input tensor. 
The kernels producing feature maps with weak activations are likely to be unimportant, as the values in the feature map with weak activations are generally small (and even zero) and have little effect on the subsequent calculations of the model~\cite{li2016pruning}. 
Inspired by this, CNNSplitter measures the importance of kernels for each class based on the feature maps.
Specifically, given $m$ samples labeled class $n$ from the training dataset, a kernel outputs $m$ feature maps. %
We calculate the sum of all values in each feature map and use the average of $m$ sums to measure the importance of the kernel for class $n$. 
Then $L$ kernels are divided into $G$ groups following the importance order. 
Consequently, a module is encoded into a bit vector $[0,1]^G$, where each bit represents whether the corresponding group of kernels is removed. 
The number of candidate modules for the $n$th class is $2^G$, and for $N$ classes, the search space size is reduced to $N \times 2^G$. %

For simplicity, if the number of kernels in a convolutional layer is less than 256, the kernels are divided into 10 groups; otherwise, they are divided into 100 groups. In this way, each kernel group has a moderate number of kernels (\eg about 10), and groups in the same convolutional layer have approximately the same number of kernels. 

\subsection{Search Strategy}
\label{subsec:searchstrategy}
A genetic algorithm~\cite{evolution2017genetic} is used to search CNN modules, which has been widely used in search-based software engineering~\cite{geneticSE_1, geneticSE_2}.
The search process starts by initializing a population of $N_I$ individuals for each of $N$ classes. 
Then, CNNSplitter performs $T$ generations, each of which consists of three operations (\ie selection, crossover, and mutation) and produces $N_I$ new individuals for each class. 
The fitness of individuals is evaluated via a performance estimation strategy that will be introduced in Section \ref{subsec:performance}. 

\subsubsection{Sensitivity-based Initialization} 
In the $0$th generation, a set of modules $M_n^0=\{m_{n, i}^0\}_{i=0}^{N_I-1}$ are initialized for class $n$,
where $n=0, 1, \dots, N-1$ and $m_{n, i}^0$ is a bit vector $[0,1]^G$. %
Two schemes are used to set the bits in each individual (\ie module): random initialization and \textit{sensitivity-based initialization}. 
Random initialization is a common scheme~\cite{nas2019survey, evolution2017genetic}. Each bit in an individual is independently sampled from a Bernoulli distribution. %
However, random initialization causes the search process to be slow or even fail.
We observed a phenomenon that some convolutional layers are sensitive to the removal of kernels, which has been also observed in network pruning ~\cite{li2016pruning}. That is, the accuracy of a CNN model dramatically drops when some particular kernels are dropped from a sensitive convolutional layer, while the loss of accuracy is not more than 0.01 when many other kernels (\eg 90\% of kernels) are dropped from an insensitive layer.

To evaluate the sensitivity of each convolutional layer, we drop out 10\% to 90\% kernels in each layer incrementally and evaluate the accuracy of the resulting model on the validation dataset. 
If the loss of accuracy is small (\eg within 0.05) when 90\% kernels in a convolutional layer are dropped, the layer is insensitive, otherwise, it is sensitive.
When initializing $m_{n, i}^0$ using sensitivity-based initialization, fewer kernel groups are dropped from the sensitive layers while more kernel groups from the insensitive layers. 
More specifically, 
a drop ratio is randomly selected from 10\% to 50\% for a sensitive layer (\ie 10\% to 50\% bit values are randomly set to 0). 
In contrast, a drop ratio is randomly selected from 50\% to 90\% for an insensitive layer.

\subsubsection{Selection, Crossover, and Mutation}
\label{subsubsec:selection}
For class $n$, to generate the population (\ie modules) of the $t$th generation, CNNSplitter performs selection, crossover, and mutation operations on the $(t-1)$th generation's population $M_n^{t-1}$.
First, the selection operation selects $N_P$ individuals from $M_n^{t-1}$ as parents according to individuals' fitness. 
Then, the single-point crossover operation generates two new individuals by exchanging part of two randomly chosen parents from $N_P$ parents. 
Next, the crossover operation iterates until $N_I$ new individuals are produced. 
Finally, the mutation operation on the $N_I$ new individuals involves flipping each bit independently with a probability $p_M$. 
For $N$ classes, selection, crossover, and mutation operations are performed in parallel, resulting in a total of $N \times N_I$ modules. 

\subsection{Performance Estimation Strategy}
\label{subsec:performance}
A module with high fitness should have the same good identification ability as the trained model $\mathcal{M}$ and only recognize the features of one specific class. 
Two evaluation metrics are used to evaluate the fitness of modules: the \textit{accuracy} and the \textit{difference} of modules.
The higher the \textit{accuracy}, the stronger the ability of the module to recognize features of the specific class. 
On the other hand, 
the greater the \textit{difference}, the more a module focuses on the specific class. 
In addition, the \textit{difference} can be used as a regularization to prevent the search from overfitting the \textit{accuracy}, as the simplest way to improve \textit{accuracy} is to allow each module to retain all the convolution kernels of $\mathcal{M}$. 
Consequently, the fitness of a module is the weighted sum of the \textit{accuracy} and the \textit{difference}. 
Furthermore, when calculating the fitness, a pruning strategy is used to improve the evaluation efficiency, making the performance estimation strategy computationally feasible.

\subsubsection{Evaluation Metrics}
\label{subsubsec:metrics}
Since a module focuses on a specific class and is equivalent to a single-class classifier, we combine modules into a composed model (CM) to evaluate them. 
That is, one module is selected from each class's $N_I$ modules, and the $N$ modules are combined into a $CM^{(N)}$ for $N$-class classification.
The $CM^{(N)}$ is evaluated on the same classification task as $\mathcal{M}$ using the dataset $D$.
The accuracy of $CM^{(N)}$ and the difference between the modules within $CM^{(N)}$ are assigned to each module.
Specifically, the accuracy and difference of each module are calculated as follows: 

\textbf{\textit{Accuracy (Acc). }}
To calculate the \textit{Acc} of $CM^{(N)}$, the $N$ modules are executed in parallel, and the output of $CM^{(N)}$ is obtained by combining modules' outputs. 
Specifically, given a $CM^{(N)}{=}\{m_n\}_{n=0}^{N-1}$, the output of module $m_n$ for the $j$th input sample labeled $class_j$ is a vector $O_{n,j}{=}[o_{n,j}^0, o_{n,j}^1, \dots, o_{n,j}^{N-1}]$, where each value corresponds to a class. Since $m_n$ is used to recognize class $n$, the $n$th value $o_{n,j}^n$ is retained. 
Consequently, the output of $CM^{(N)}$ is $O_j{=}[o_{0,j}^0, o_{1,j}^1 \dots, o_{N-1,j}^{N-1}]$, and the \textit{Acc} of $CM^{(N)}$ is calculated as follows:
\begin{gather}
    Acc = \frac{1}{|D|} \sum_j^{|D|} pred(j), \\
    pred(j) = 
    \begin{cases}
    1, & \mbox{if }\mathop{\arg\max}\limits_{n=0, 1, \dots, N-1}O_j = class_j \\
    0, & \mbox{if }\mathop{\arg\max}\limits_{n=0, 1, \dots, N-1}O_j \ne class_j .
    \end{cases}
\end{gather} 

\textbf{\textit{Difference (Diff). }}%
Since a module can be regarded as a set of convolution kernels, the difference between two modules can be measured by the Jaccard Distance (\textit{JD}) that measures the dissimilarity between two sets. 
The \textit{JD} between set $A$ and set $B$ is obtained by dividing the difference of the sizes of the union and the intersection of two sets by the size of the union: 
\begin{gather}
    JD(A, B) = \frac{|A \cup B| - |A \cap B|}{|A \cup B|}. \label{eq:jaccard}
\end{gather}
If the $JD(A,B)=1$, there is no commonality between set $A$ and set $B$, and if it is 0, then they are exactly the same.
Based on \textit{JD}, the \textit{Diff} value of $CM^{(N)}$ is the average value of \textit{JD} between all modules:
\begin{gather}
    Diff = \frac{2}{N \times (N-1)} \times \sum_{0 \leq i < j \leq N-1} JD(m_i, m_j). \label{eq:diff}
\end{gather}
Based on \textit{Acc} and \textit{Diff}, the fitness value of $CM^{(N)}$ is calculated via:
\begin{gather}
    fitness = \alpha \times Acc + (1-\alpha) \times Diff, \label{eq:fitness}
\end{gather}
where $\alpha$ is a weighting factor and $0 < \alpha < 1$. In practice, $\alpha$ is set to a high value (\eg 0.9) because high accuracy is a prerequisite for the availability of modules.
The fitness value of $CM^{(N)}$ is then assigned to the $N$ modules within $CM^{(N)}$. 
Since each module is used in multiple CMs, a set of fitness values is assigned to each module. 
The maximum value of the set is a module's final fitness. 

\subsubsection{Decode}
\label{subsubsec:decode}
\begin{figure}
    \centering
    \includegraphics[width=8cm]{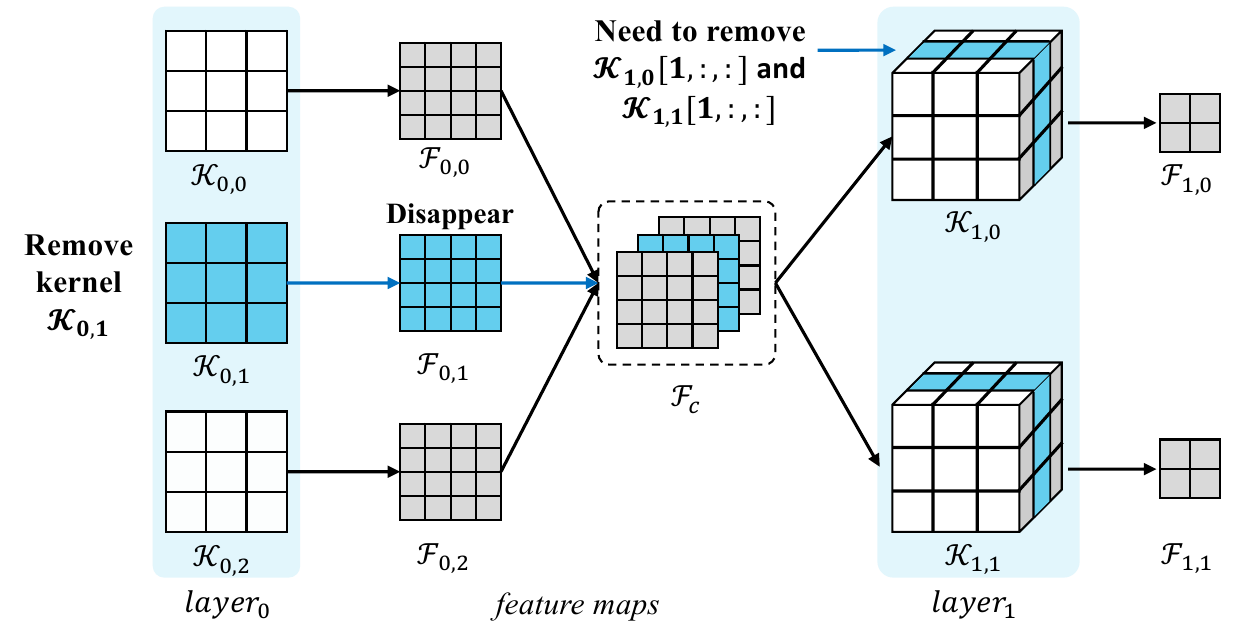}
    \caption{The process of removing convolution kernels.}
    \label{fig:decode}
    \vspace{-6pt}
\end{figure}

To evaluate modules, each bit vector is transformed into a runnable module by removing the kernel groups from $\mathcal{M}$ corresponding to the bits of value 0.
Since removing kernels from a convolutional layer affects the convolutional operation in the later convolutional layer, the kernels in the latter convolutional layer need to be modified to ensure that the module is runnable. 

Figure \ref{fig:decode} shows the process of removing convolution kernels. 
During the convolution, the three kernels $k_{0,*} \in \mathbb{R}^{3 \times 3}$ in $layer_0$ output three feature maps $F_{0,*} \in \mathbb{R}^{4 \times 4}$ that are then combined in a feature map $F_c \in \mathbb{R}^{3 \times 4 \times 4}$ and fed to $layer_1$. 
By sliding on $F_c$, kernels $k_{1,*} \in \mathbb{R}^{3 \times 3 \times 3}$ in $layer_1$ perform the convolution and output two feature maps $F_{1,*} \in \mathbb{R}^{2 \times 2}$. 
If $k_{0,1}$ in $layer_0$ is removed, the feature map $F_{0,1}$ generated by $k_{0,1}$ is also removed. 
The input of $layer_1$ becomes a different feature map $F^{'}_c \in \mathbb{R}^{2 \times 4 \times 4}$, the dimension of which does not match that of $k_{1,*} \in \mathbb{R}^{3 \times 3 \times 3}$ in $layer_1$, causing the convolution to fail. 

To solve the dimension mismatch problem, we remove the part of $k_{1,*}$ that corresponds to $F_{0,1}$, ensuring the first dimension of $k_{1,*}$ to match with that of $F^{'}_c$. 
For instance, since $F_{0,1}$ is removed, $k_{1,0}[1,:,:]$, which performs convolution on $F_{0,1}$, becomes redundant and causes the dimension mismatch. 
We remove $k_{1,0}[1,:,:]$ and the transformed kernel $k^{'}_{1,0} \in \mathbb{R}^{2 \times 3 \times 3}$ can perform convolution on $F^{'}_c$. 

In addition, since the residual connection adds up the feature maps output by two convolutional layers, the number of kernels removed from the two convolutional layers must be the same to ensure that the output feature maps match in dimension. 
When constructing a bit vector, we treat the two convolutional layers as one layer and use the same segment to represent the two layers so that they always remove the same number of kernels. 

\subsubsection{Pruning-based Evaluation}

Since the fitness of each module comes from the one with the highest fitness among the CMs the module participates in, the number of $CM^{(N)}$ that CNNSplitter needs to evaluate is $(N_I)^N$.
The time complexity is $O(n^N)$, which could be too high to finish the evaluation in a limited time. 
To reduce the overhead, a pruning strategy is designed, which is based on the following fact: if the accuracy of $CM^{(N)}$ is high, the accuracy of the $CM^{(n)}$ (\eg $CM^{(2)}$ for the binary classification) composed of the modules within the $CM^{(N)}$ is also high. 
If the accuracy of a module is low, the accuracy of the $CM^{(n)}$ containing the module is also lower than the $CM^{(n)}$ containing modules with high accuracy.
In addition, the number of $CM^{(n)}$ is much smaller than that of $CM^{(N)}$. 
For instance, the $N$-class classification task can be decomposed into $N/2$ binary classification subtasks, resulting in $N/2 \times (N_I)^2$ $CM^{(2)}$. 

Therefore, the $N$-class classification task is decomposed into several subtasks. The accuracy of $CM^{(n)}$ is evaluated, and the top $N_{top}$ $CM^{(n)}$ with high accuracy are selected to be combined into $CM^{(N)}$.
Through continuous evaluation, selection, and composition, a total of $(N_{top})^2$ $CM^{(N)}$ are composed. 
The time complexity is $O(n^2)$, which is lower than the original time complexity $O(n^N)$.

\section{\projectName: Gradient-based Compressed Modularization}
\label{sec:approach_decom}
\begin{figure}[t]
	\centering
	\includegraphics[width=8.8cm]{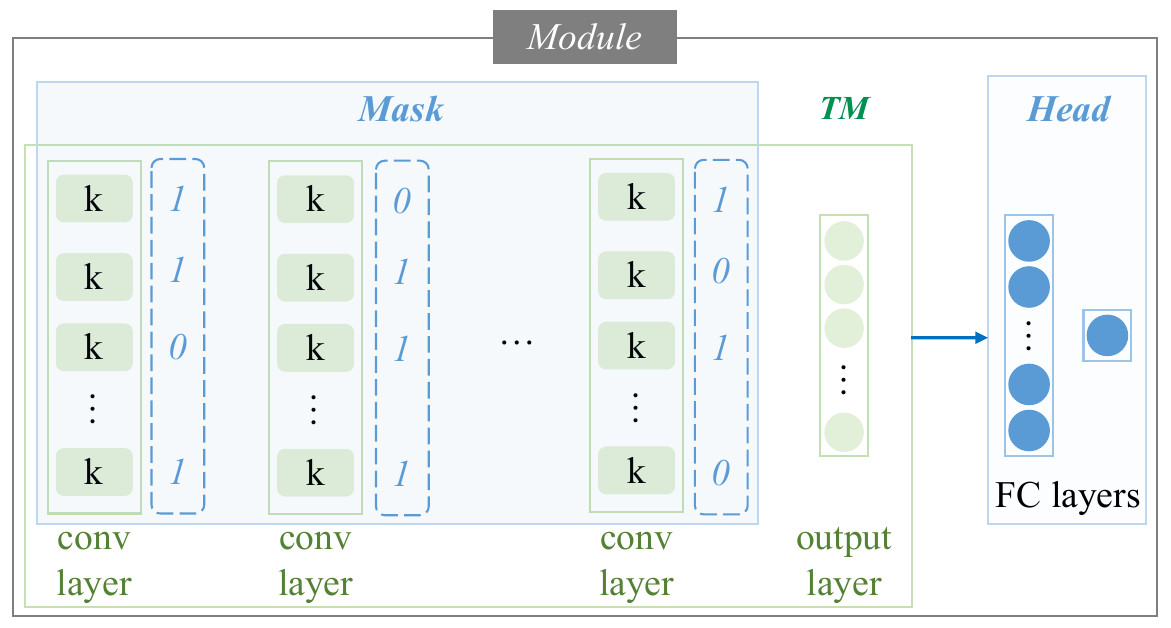}
    \caption{An illustration of creating a module through $mask \odot TM \oplus head$.}
    \label{fig:module}
    \vspace{-6pt}
\end{figure}

We propose a gradient-based compressed modularization approach, named \projectName, to decompose a trained CNN model \textit{TM} into smaller modules.
To achieve the modularization of a \textit{TM}, a \textit{mask} and a \textit{head} are used to create a module through $mask \odot TM \oplus head$. 
Figure \ref{fig:module} illustrates the creation of a module.
A \textit{mask} consists of an array of 0s and 1s, where 0 (or 1) indicates that the corresponding convolution kernels in the \textit{TM} are removed (or retained). 
The operation $mask \odot TM$ removes kernels from \textit{TM}, resulting in a \textit{masked TM}.
A \textit{head} consists of FC layers and converts an $N$-dimensional input into a $1$-dimensional output.
The operation $\oplus head$ appends a \textit{head} after the \textit{masked TM}, resulting in a module. 
As a result, the constructed module is a binary classifier. 
The output value of a module greater than 0.5 indicates that the input belongs to the target class, and vice versa.

Subsequently, the decomposition is a training process for \textit{masks} and \textit{heads}. 
As shown in Algorithm \ref{algo:overall}, the training process mainly consists of the forward propagation (Line \ref{algo:overall:forward}) and the backward propagation (Line \ref{algo:overall:backward}). 
The forward propagation computes the prediction of the composed model (\textit{CM}) constructed using $N$ modules. The $N$ modules are created based on the current \textit{masks} and \textit{heads}.
The backward propagation optimizes \textit{masks} and \textit{heads} using the gradient descent based on the current prediction.
Section \ref{subsec:mask} and Section \ref{subsec:head} provide detailed descriptions of the \textit{masks} and \textit{heads}, respectively.
Section \ref{subsec:optim} explains how to optimize \textit{masks} and \textit{heads} to obtain modules and achieve CNN model decomposition.

\begin{algorithm}[t]
\small
\caption{Overall \projectName Algorithm}
\label{algo:overall}
\LinesNumbered
\SetKwFunction{Forward}{Forward}
\SetKwFunction{Backward}{Backward}
\SetKwFunction{Bin}{Bin}
\KwIn{Training dataset $D$ and trained CNN model $TM$.}
\KwOut{$N$ binarized \textit{masks} and $N$ \textit{heads}.}
\BlankLine
$S_{m} = \{mask_i\}_{i=1}^N$, where $mask_i[:]>0 $\; \label{algo:overall:initmask}
$S_{h} = \{head_i\}_{i=1}^N$, where $head_i$ consists of FC layers\; \label{algo:overall:init_head}
$results = []$\;
\For{$e = 1, 2, ..., E$}{
    \For{$input$, $label$ in $D$}{ 
        \tcp{Forward() is shown in Alg.\ref{algo:forward}}
        $pred=$ \Forward{$S_{m}$, $S_{h}$, $input$} \; \label{algo:overall:forward}
        \tcp{Backward() is shown in Alg.\ref{algo:backward}}
        $S_{m},S_{h}{=}$ \Backward{$S_{m}$, $S_{h}$, $pred$, $label$, $e$} \; \label{algo:overall:backward}
        $results.append([$ \Bin{$S_m$}, $S_h])$\;
    }
}
select $S_m$ and $S_h$ from $results$\; \label{algo:overall:select}
\Return $S_{m}$, $S_{h}$
\end{algorithm}

\begin{algorithm}[t]
\small
\caption{Forward propagation}
\label{algo:forward}
\LinesNumbered
\SetKwFunction{Bin}{Bin}
\SetKwFunction{Concat}{Concat}
\KwIn{$N$ masks $S_m$, $N$ heads $S_h$, and input data $input$.}
\KwOut{prediction $predict$.}
\BlankLine
$modules\_pred=$ []\; 
\For{$i=1, 2, ..., N$} 
{\tcp{compute each module's prediction}
    $mask_i^b =$ \Bin{$S_m[i]$}\;  \label{algo:forward:binarize}
    $out = mask_i^b \odot TM(input)$\;  \label{algo:forward:mask}
    $head_i=S_h[i]$\;
    $pred = head_i(out)$\; \label{algo:forward:head}
    $modules\_pred.append(pred)$\;  
}
$predict =$ \Concat{$modules\_pred$}\; \label{algo:forward:concat}
\Return $predict$
\end{algorithm}

\subsection{Mask}
\label{subsec:mask}
A \textit{mask} should consist of binarized values (\ie 0s and 1s) to indicate which convolution kernels in the \textit{TM} are removed (or retained). 
On the other hand, during the training process, the values in a \textit{mask} should be continuous numerical values, instead of binarized values, to enable gradient descent. 
Therefore, the \textit{mask} is initialized with random positive numbers during the training process (Line \ref{algo:overall:initmask} in Algorithm \ref{algo:overall}). 
\projectName uses a binarization function to transform the \textit{mask}, resulting in a binarized mask as an intermediate value (Line \ref{algo:forward:binarize} in Algorithm \ref{algo:forward}).
In order to achieve the transformation, the %
binarization function $Bin$ is defined as follows: 
\begin{gather}
    x_b = Bin(x) =
    \begin{cases} 
    1, & \mbox{if } x\mbox{ \textgreater \ 0}, \\ 
    0, & \mbox{otherwise}, \\ 
    \end{cases} \label{eq:bin}
\end{gather}
where $x_b$ is the binarized variable and $x$ is the real-valued variable.

To achieve the effect of removing convolution kernels according to the binarized mask in forward propagation, we multiply the output of each convolution kernel by the corresponding value in the binarized mask (Line~\ref{algo:forward:mask} in Algorithm~\ref{algo:forward}). 
For instance, as shown in Figure \ref{fig:decode}, the feature map $F_{0,1}$ generated by the convolution kernel $k_{0,1}$ is multiplied by the corresponding value 0 in the binarized mask, resulting in all values in $F_{0,1}$ being 0. 
After multiplying the outputs of the previous convolutional layer's kernels by the corresponding values in the binarized mask, the outputs of the convolution kernels that should be removed are set to 0s.
In this way, the convolution kernels that should be removed will not affect the subsequent prediction, thus enabling the simulation of removing convolution kernels.
Note that, according to the trained mask, the unwanted convolution kernels will be removed from the modules during module reuse.

\subsection{Head}
\label{subsec:head}
A masked \textit{TM} cannot be used as a module directly, as the output of the masked \textit{TM} is still $N$-dimensional. 
Although the $i$th value in the output could indicate the probability that the input belongs to the $i$th class, using the $i$th value in the output as the prediction of a module is problematic.
The output layer of the \textit{TM} predicts based on the features extracted by all convolution kernels. 
There could be significant bias in the prediction of masked \textit{TM}, as the output layer predicts based on only retained convolution kernels.
For instance, regardless of which class the input belongs to, the $i$th value in the output of a masked \textit{TM} that recognizes the target class $i$ is always larger than other values in the output.
As a result, a masked \textit{TM} that recognizes target class $i$ always %
classifies the inputs to the target class $i$ even if the inputs actually belong to other classes.

Therefore, an additional output layer (\ie \textit{head}) consisting of two FC layers and a $sigmoid$ activation function is appended after the masked $TM$ (see Figure \ref{fig:module}). The numbers of neurons in the two FC layers are $N$ and $1$, respectively. 
The \textit{head} transforms the $N$-class prediction of masked $TM$ to the binary classification prediction, resulting in the prediction of a module (Line~\ref{algo:forward:head} of Algorithm~\ref{algo:forward}). 
As a result, each module recognizes a target class and estimates the probability of an input belonging to the target class. 
Since each module recognizes a target class and outputs a probability value between 0 and 1, $N$ modules can be aggregated as a composed model for the $N$-class classification task (Line \ref{algo:forward:concat} of Algorithm~\ref{algo:forward}). 

\begin{algorithm}[t]
\small
\caption{Backward propagation}
\label{algo:backward}
\LinesNumbered
\SetKwFunction{CrossEntropy}{CrossEntropy}
\SetKwFunction{PercentKernels}{PercentKernels}
\SetKwFunction{GradientDescent}{GradientDescent}
\KwIn{$N$ masks $S_m$, $N$ heads $S_h$, prediction $pred$, data labels $label$, and the current epoch $e$.}
\KwOut{Updated masks $S_m$ and heads $S_h$.}
\BlankLine

$actions=[0, 1]^E$\;  \label{algo:backward:init_action}

\uIf{$actions[e] == 0$}{ \label{algo:backward:action_1}
    $update\_obj = [S_h]$\;
    $weight = 0$\;
}
\Else{
  $update\_obj = [S_m, S_h]$\;
  $weight = \beta$\; \label{algo:backward:action_2}
}

$loss_1 =$ \CrossEntropy{$pred$, $label$}\;  \label{algo:backward:loss1}
$loss_2 =$ \PercentKernels{$S_m$}\; \label{algo:backward:loss2}
$loss = loss_1 + weight \times loss_2$\; \label{algo:backward:loss}
\GradientDescent{$loss$, $update\_obj$}\; \label{algo:backward:backprop}

\Return{$S_m$, $S_h$}
\end{algorithm}

\subsection{Optimization by Gradient Descent}
\label{subsec:optim}
The goal of modularization is to obtain $N$ \textit{masks} and $N$ \textit{heads} to decompose a \textit{TM} into $N$ modules.
Each module can recognize a target class and should retain only the convolution kernels necessary for recognizing the target class. 
To achieve the goal, \projectName needs to optimize \textit{masks} to remove the redundant convolution kernels as many as possible and optimize \textit{heads} to predict based on the retained convolution kernels.
The optimization is achieved by minimizing weighted loss through gradient descent-based backward propagation, which is shown in Algorithm~\ref{algo:backward}.

Specifically, to ensure that each module can recognize the target class well, $loss_1$ is defined as the cross entropy between the prediction of \textit{CM} and the actual label (Line \ref{algo:backward:loss1}). 
By minimizing $loss_1$, the prediction performance of the \textit{CM} is improved, while the improvement of \textit{CM} is essentially owing to the improvement of modules in recognizing target classes.
The \textit{masks} tend to make more values greater than zero to retain more convolution kernels,
and the \textit{heads} are trained to predict based on the retained convolution kernels.
To constraint modules to retain only the necessary kernels, $loss_2$ is defined as the percentage of retained convolution kernels (Line \ref{algo:backward:loss2}).
By minimizing $loss_2$, the \textit{masks} tend to have fewer values that are greater than zero, resulting in fewer convolution kernels.

We define $loss$ as the weighted sum of $loss_1$ and $loss_2$ (Line \ref{algo:backward:loss}).
By minimizing $loss$, the \textit{masks} are optimized to have as few necessary values greater than zero as possible, \ie retain only the necessary convolution kernels.
Meanwhile, the \textit{heads} are trained to predict based on the retained convolution kernels.
The larger the value of $weight$, the more the convolution kernels \projectName tends to remove. 
In our experiments, the value of $weight$ is usually small (\eg 0.1), allowing \projectName to remove convolution kernels carefully, thus avoiding much impact on the recognition ability of modules. %

When minimizing $loss$, one problem is that the poor predictions of modules cannot guide the optimization well in the early stages of optimization, as the \textit{heads} are initialized randomly. 
Moreover, 
$loss_2$ could affect the optimization of \textit{heads}, leading to the increased loss of accuracy.
Therefore, a strategy is designed for the optimization, which is shown in Line 
\ref{algo:backward:init_action} to \ref{algo:backward:action_2} in Algorithm \ref{algo:backward}.
$actions$ is a bit vector $[0,1]^E$ (Line \ref{algo:backward:init_action}), which is used to select the optimization object in an epoch $e$.
When the value of $actions[e]$ is 0, $weight$ is set to zero, and \projectName minimizes only $loss_1$ to optimize \textit{heads}.
When the value of $actions[e]$ is 1, $weight$ is set to $\beta$, and \projectName minimizes $loss$ to optimize \textit{masks} and \textit{heads} jointly.
With the strategy, \projectName can optimize only \textit{heads} in the first few epochs to recover the loss of accuracy caused by the randomly initialized \textit{heads}.
On the other hand, \projectName can minimize only $loss_1$ after every several epochs of minimizing $loss$ to recover the loss of accuracy caused by the removal of convolution kernels.

The gradient descent-based optimization is applied to minimize $loss$ (Line \ref{algo:backward:backprop}). When minimizing $loss$ by gradient descent, it is important to note that the common backward propagation based on gradient descent cannot be directly applied to update \textit{masks}, as the derivative of the $Bin$ function is zero almost everywhere.
Fortunately, the technique called straight-through estimator (STE)~\cite{ste} has been proposed to address the gradient problem occurring when training neuron networks with binarization function (\eg \textit{sign})~\cite{ste, hubara2016binarized,binarynn}. The function of STE is defined as follows: 
\begin{gather}
    clip(x, -1, 1) = max(-1, min(1, x)),
\end{gather}
where $x$ is the gradient value calculated in the previous layer and used to estimate the gradient in current layer.
In our work, based on STE, the gradient of $Bin$ (defined in Equation \ref{eq:bin}) is calculated as follows:
\begin{gather}
    \frac{\partial loss}{\partial x} = clip(\frac{\partial loss}{\partial x_b}, -1, 1).
\end{gather}

\section{Applications of CNN modularization}
In this section, we present two applications of CNNSplitter and \projectName: 1) patching weak CNN models and 2) developing new models via composition.
\subsection{Application 1: Patching Weak CNN Models through Modularization and Composition}
\label{subsec:patching}
\begin{figure}[!t]
    \centering
    \includegraphics[width=8.3cm]{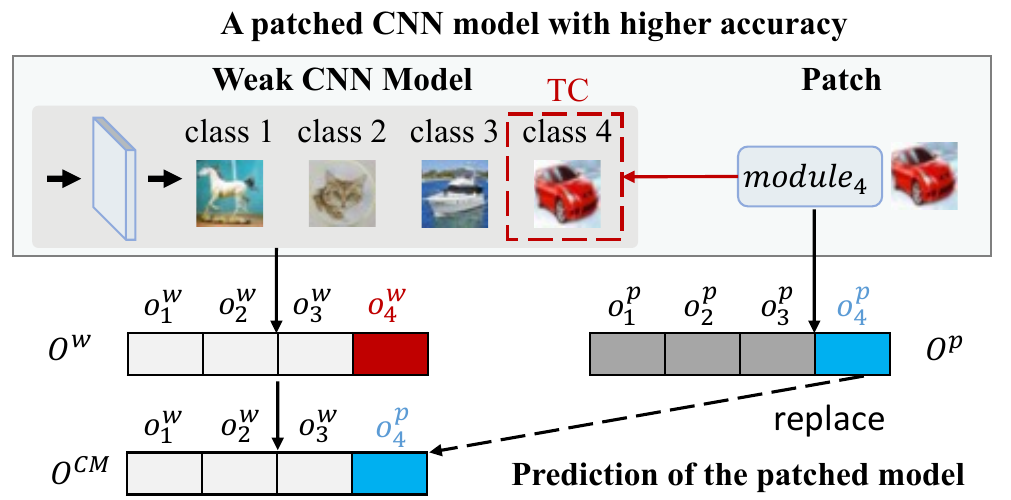}
    \caption{Patching a weak CNN model.}
    \label{fig:patching}
    \vspace{-12pt}
\end{figure}

The weak CNN model can be improved by patching the target class (TC).
To identify the TC of a weak CNN model, developers can use test data to evaluate the weak CNN model's classification performance (\eg precision and recall) of each class. The class in which the weak CNN model achieves poor classification performance is regarded as TC. 
As illustrated in Figure \ref{fig:patching}, 
the TC is replaced with the corresponding module from a strong model. 
To find the corresponding module, a developer can evaluate the accuracy of a candidate model on TC. 
If the candidate model's accuracy exceeds that of the weak model, its module can be used as a patch. 

Formally, 
given a weak CNN model $\mathcal{M}_w$, suppose there exists a strong CNN model $\mathcal{M}_s$ whose classification task intersects with that of $\mathcal{M}_w$. For instance, both $\mathcal{M}_w$ and $\mathcal{M}_s$ can recognize TC $n$. Then, the corresponding module $m_n$ from $\mathcal{M}_s$ can be used as a patch to improve the ability of $\mathcal{M}_w$ to recognize TC $n$. 
Specifically, $\mathcal{M}_w$ and $m_n$ are composed into a CM that is the patched CNN model. 
Given an input, $\mathcal{M}_w$ and $m_n$ run in parallel and the outputs of them are $O^w=[o^w_0, o^w_1, \dots, o^w_{N-1}]$ and $O^p=[o^p_0, o^p_1, \dots, o^p_{N-1}]$, respectively. 
Then, the output $O^{CM}$ of CM is obtained by replacing the prediction corresponding to TC $n$ of $O^w$ with that of $O^p$. 

A straightforward way is to directly replace $o^w_n$ with $o^p_n$, resulting in $O^{CM}=[o^w_0, \dots, o^p_n, \dots, o^w_{N-1}]$. The index of the maximum value in $O^{CM}$ is the predicted class. 
However, the comparison between $o^p_n$ and the other values in $O^{CM}$ is problematic: since $\mathcal{M}_w$ and $\mathcal{M}_s$ are different models that are trained on the different datasets or have different network structures, there could be significant differences in the distribution between the outputs of $\mathcal{M}_w$ and $\mathcal{M}_s$. 
For instance, we have observed that the output values of a model could be always greater than that of the other one, resulting in the outputs of a module decomposed from the strong model being always larger/smaller than the outputs of a weak model. 
This problem could cause error prediction when calculating the prediction of CM; thus, $O^w$ and $O^p$ are normalized before the replacement. 
Specifically, since the outputs on the training set can reflect the output distribution of a module, we collect the outputs of $m_n$ on the training data with the class label $n$. 
Then, the minimum and maximum values of the output's distribution can be estimated using the collected outputs.  
For instance, ($min$, $max$) are the minimum and maximum values of the collected outputs of $m_n$. 
The normalized $o^p_n$ is $\frac{o^p_n-min}{max - min}$. 
In addition, the $softmax$ is used over $O^w$ to scale the values in $O^w$ between 0 and 1. 
Finally, the prediction of CM is obtained by replacing $o^w_n$ in normalized $O^w$ with normalized $o^p_n$.

\subsection{Application 2: Developing CNN Models via Composition}
\label{sec:approach_reuse}
When a developer needs an $N$-class classification CNN model, 
the existing trained CNN models (\textit{TMs}) shared by third-party developers could be decomposed and reused.
Developing a %
new CNN model (\textit{CM}) %
by composing reusable modules %
consists of three steps: \textit{module creation}, \textit{module evaluation}, and \textit{module reuse}.
\textit{Module creation} mainly involves the removal of irrelevant convolution kernels from models according to masks, which is similar to the "Decode" operation in CNNSplitter introduced in Section \ref{subsubsec:decode}. The following sections will introduce \textit{module evaluation} and \textit{module reuse}.

\begin{figure}
    \centering
    \includegraphics[width=0.7\columnwidth]{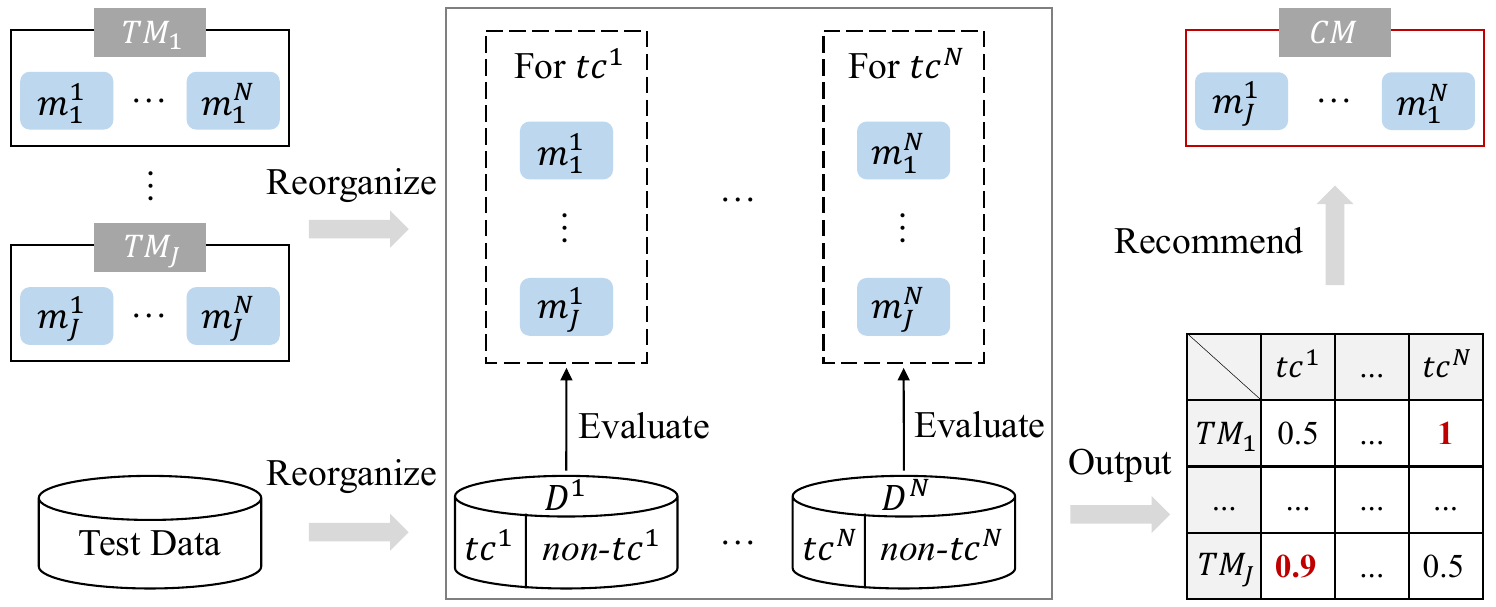}
    \caption{The process of evaluating modules.}
    \label{fig:evaluate_modules}
    \vspace{-6pt}
\end{figure}

\subsubsection{Module Evaluation}
\label{subsec:module_evaluate}
To obtain a more accurate \textit{CM} than \textit{TMs} or a new \textit{CM} that can achieve %
comparable performance to the model trained from scratch, module evaluation is a key step, which can identify the module with the best recognition ability for each target class. 

As shown in Figure \ref{fig:evaluate_modules}, given a set of trained CNN models $\{TM_j\}_{j=1}^J$ for $N$-class classification and test data for $N$-class classification, \projectName first reorganizes them according to target classes. 
Specifically, for each target class $tc^n$, $J$ corresponding modules that can recognize the target class $tc^n$ are put together to form a set of candidate modules $\{m_j^n\}_{j=1}^J$. 
As a result, there are $N$ sets of candidate modules.
For each set, the module with the best recognition ability is recommended to the developer. 
Since each module $m_j^n$ is a binary classification model that recognizes whether an input belongs to $tc^n$, the candidate modules from the same set can be tested and compared on the same binary classification task. 
As $N$ sets of candidate modules correspond to $N$ binary classification tasks, the test data for $N$-class classification needs to be reorganized to form $N$ data sets $\{D^n\}_{n=1}^N$ for binary classification, each with the target class $tc^n$ as the positive class and the other $N{-}1$ non-target classes \textit{non-}$tc^n$ as the negative class. 
Considering that $D^n$ could be imbalanced due to the disproportion among the number of samples of positive class and negative class, F1-score is used to measure the recognition ability of $m_j^n$ for $tc^n$. 
For each target class $tc^n$, the module with the highest F1-score among the candidate modules $\{m_j^n\}_{j=1}^J$ is recommended to the developer. 

\subsubsection{Module Reuse}
In the module composition, $N$ recommended modules are combined into a \textit{CM} for $N$-class classification. 
The composition is simple, and the \textit{CM} runs like a common CNN model. 
Specifically, given an input, $N$ modules run in parallel, and their outputs are concatenated as the output of the \textit{CM}. 
The index of the maximum value in the output is the final prediction. 
Since each module in the \textit{CM} has the best recognition ability for the corresponding target class, the \textit{CM} can outperform any of the $J$ trained CNN models or achieve competitive performance in accuracy compared to the model trained from scratch.

\section{Experiments}
\label{sec:experiments}

To evaluate the effectiveness of the proposed approaches, in this section, we first introduce the benchmarks and experimental setup and then discuss the experimental results.
Specifically, we evaluate CNNSplitter and \projectName by answering the following research questions:

\begin{itemize}[leftmargin=*]
\item RQ1: How effective are the proposed techniques in modularizing CNN models?
\item RQ2: Can the recognition ability of a weak model for a target class be improved by patching?
\item RQ3: Can a composed model, built entirely by combining modules, outperform the best trained model?
\item RQ4: Can a CNN model for a new task be built through modularization and composition while maintaining an acceptable level of accuracy?
\item RQ5: How efficient is \projectName in modularizing CNN models and how efficient is the composed CNN model in prediction?
\end{itemize}

\subsection{Benchmarks}
\paragraph{1) Datasets}
We evaluate the proposed techniques on the following three datasets, which are widely used for evaluation in related work~\cite{nnmodularity2022icse,icse21discriminiate,feng2020deepgini}. %

\textbf{CIFAR-10.} The CIFAR-10 dataset~\cite{cifar10} contains natural images with resolution $32\times32$, which are drawn from 10 classes including airplanes, cars, birds, cats, deer, dogs, frogs, horses, ships, and trucks. 
The initial training dataset and testing dataset contain 50,000 and 10,000 images, respectively. 

\textbf{CIFAR-100.} %
CIFAR-100~\cite{cifar10} consists of $32\times32$ natural images in 100 classes, with 500 training images and 100 testing images per class.  

\textbf{SVHN.} The Street View House Number (SVHN) dataset ~\cite{svhn} contains colored digit images 0 to 9 with resolution $32\times32$. 
The training and testing datasets contain 604,388 and 26,032 images, respectively.

\paragraph{2) Models}
We evaluate the proposed techniques on the following three typical CNN structures, which are widely used in popular networks ~\cite{lenet, alexnet, vgg, resnet, googlenet, inception}.

\textbf{SimCNN.} SimCNN represents a class of CNN models with a basic structure, such as LeNet~\cite{lenet}, AlexNet~\cite{alexnet}, and VGGNet~\cite{vgg}, essentially constructed by stacking convolutional layers. The output of each convolutional layer can only flow through each layer in sequential order. 
Without loss of generality, the SimCNN in our experiments is set to contain 13 convolutional layers and 3 FC layers, totaling 4,224 convolution kernels.

\textbf{ResCNN.} ResCNN represents a class of CNN models with a complex structure, such as ResNet~\cite{resnet}, WRN~\cite{wrn}, and MobileNetV2~\cite{mobilenetv2}, constructed by convolutional layers and residual connections. A residual connection can go across one or more convolutional layers, allowing the output of a layer not only to flow through each layer in sequential order but also to be able to connect with any following layer.
Without loss of generality, the ResCNN in our experiments is set to have 12 convolutional layers, 1 FC layer, and 3 residual connections, totaling 4,288 convolution kernels.

\textbf{InceCNN.} InceCNN represents a class of CNN models with a complex structure, such as GoogLeNet~\cite{googlenet} and Inception-V3~\cite{szegedy2016rethinking}, constructed by branched convolutional layers. The branched convolutional layers mean that the outputs of several convolutional layers are concatenated as one input to be fed into the next branched convolutional layers. 
Without loss of generality, the InceCNN in our experiments is set to have 12 convolutional layers, 1 FC layer, and 3 branched layers, totaling 3,200 convolution kernels.

All the experiments are conducted on Ubuntu 20.04 server with 64 cores of 2.3GHz CPU, 128GB RAM, and NVIDIA Ampere A100 GPUs with 40 GB memory. 

\begin{figure}
    \centering
    \includegraphics[width=9cm]{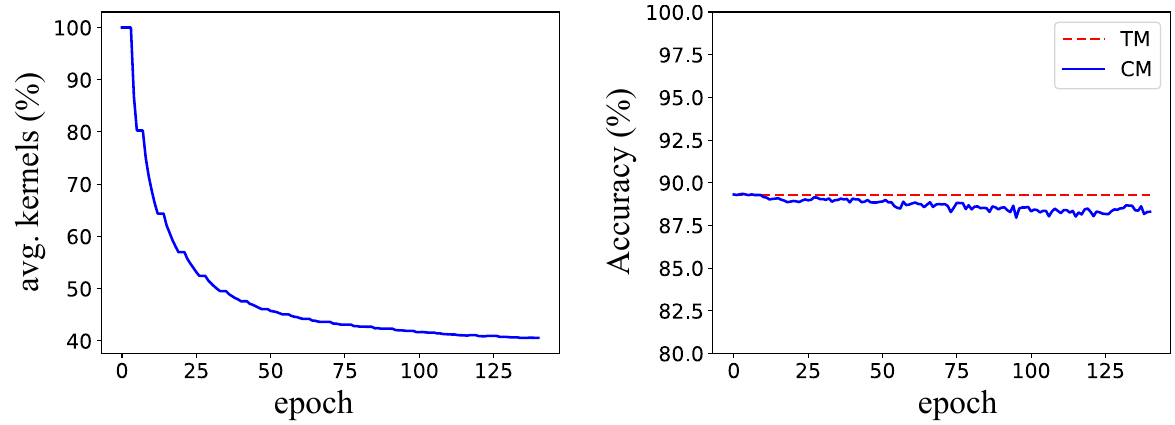}
    \caption{The convergence process of GradSplitter, including the average percentage of retained kernels in a module (the left sub-figure) and the validation accuracy of the \textit{TM}/\textit{CM} (the right sub-figure) during modularization. 
    }
    \label{fig:module_log}
    \vspace{-6pt}
\end{figure}

\subsection{Experimental Results}
\label{subsec:exp_result}
\noindent\textbf{\textit{RQ1: How effective are the proposed techniques in modularizing CNN models?}}
\paragraph{1) Setup.} 
\textbf{Training settings.}
To answer RQ1, 
SimCNN and ResCNN are trained on CIFAR10 and SVHN, resulting in four strong CNN models: SimCNN-CIFAR, SimCNN-SVHN, ResCNN-CIFAR, and ResCNN-SVHN. 
The training datasets of CIFAR10 and SVHN are divided into two parts in the ratio of $8{:}2$, respectively. The 80\% samples are used as the training set while the 20\% samples are used as the validation set.
On both CIFAR10 and SVHN datasets, SimCNN and ResCNN are trained with mini-batch size 128 for 200 epochs. The initial learning rate is set to 0.01 and 0.1 for SimCNN and ResCNN, respectively, and the initial learning rate is divided by 10 at the 60th and 120th epoch for SimCNN and ResCNN respectively. 
All the models are trained using data augmentation~\cite{dataaugmentation} and SGD with a weight decay~\cite{weightdecay} of $10^{-4}$ and a Nesterov momentum ~\cite{sutskever2013importance} of 0.9.
After completing the training, the trained models are evaluated on testing datasets.

\textbf{Modularization settings.}
CNNSplitter applies a genetic algorithm to search CNN modules, following the common practice~\cite{genetic2019, suganuma2018exploiting, real2017large}, the number of individuals $N_I$ and the number of parents $N_P$ in each generation are set to 100 and 50, respectively.
The mutation probability $p_M$ is generally small~\cite{genetic2019, suganuma2018exploiting} and is set to $0.1$. 
The weighting factor $\alpha$ is set to $0.9$. 
For the sake of time, an early stopping strategy ~\cite{q-learning-earlystopping, goodfellow2016deep} is applied, and the maximum number of generations is set as $T=200$. 
A trained CNN model $\mathcal{M}$ is modularized with reference to the validation set, which was not used in model training. 
After completing the modularization, the resulting modules are evaluated on the testing dataset.

\projectName initializes \textit{masks} by filling positive values and initialize \textit{heads} randomly. 
When initializing $actions$ (Line \ref{algo:backward:init_action} in Algorithm \ref{algo:backward}), we set $actions[1{:}5]{=}[0]{\times}5$ and $actions[6{:}E]{=}[1,1,1,1,1,0,0] \times \frac{E-5}{7}$. 
$actions[e]{=}0$ indicates that \projectName trains only the \textit{heads} in the epoch $e$.
$actions[e]{=}1$ indicates that \projectName jointly trains both \textit{masks} and \textit{heads} in the epoch $e$.
As a result, \projectName trains only the \textit{heads} in the first 5 epochs. 
Then, \projectName trains only the \textit{heads} for 2 epochs after every 5 epochs of joint training.
The training process iterates 145 epochs (\ie $E{=}145$) with a learning rate of 0.001.
By default, $\beta$ in the weighted sum of $loss_1$ and $loss_2$ is set to 0.1. 
We also investigate the impact of $\beta$ on modularization in Section \ref{subsec:exp_result}.

\begin{table}[t]
\setlength\tabcolsep{2.4pt}
\caption{The modularization results on four strong models. 
} 
\vspace{-6pt}
\label{tab:module_results_strong}
\scriptsize
\begin{center}
{
\begin{tabular}{ccccccccc}
\toprule
\multirow{2}{*}{\textbf{Model}} & \multicolumn{2}{c}{\textbf{TM}}        & \multicolumn{3}{c}{\textbf{Accuracy of a CM}}                                           & \multicolumn{3}{c}{\textbf{Average percentage of kernels in a module}}              \\ \cmidrule(lr){2-3} \cmidrule(lr){4-6} \cmidrule(lr){7-9}  
                                & \textbf{Acc.}     & \textbf{\# Kernels} & \textbf{CNNSplitter}       & \textbf{GradSplitter}       & \textbf{Increment} & \textbf{CNNSplitter} & \textbf{GradSplitter} & \textbf{Reduction} \\ \midrule \midrule
SimCNN-CIFAR10                  & 89.77\%          & 4224                & 86.07\% (-3.70\%)          & 88.90\% (-0.87\%)           & 2.83\%             & 61.96\%              & 41.22\%               & 20.74\%            \\
SimCNN-SVHN                     & 95.41\%          & 4224                & 93.85\% (-1.56\%)          & 95.67\% (+0.26\%)           & 1.82\%             & 52.79\%              & 35.18\%               & 17.61\%            \\
ResCNN-CIFAR10                  & 90.41\%          & 4288                & 85.64\% (-4.77\%)          & 89.08\% (-1.33\%)           & 3.44\%             & 58.26\%              & 40.63\%               & 17.63\%            \\
ResCNN-SVHN                     & 95.06\%          & 4288                & 93.52\% (-1.54\%)          & 94.70\% (-0.36\%)           & 1.18\%             & 54.03\%              & 30.48\%               & 23.55\%            \\ \midrule
\textbf{Average}                & \textbf{92.66\%} & \textbf{4256}       & \textbf{89.77\% (-2.89\%)} & \textbf{92.09\%  (-0.58\%)} & \textbf{2.32\%}    & \textbf{56.76\%}     & \textbf{36.88\%}      & \textbf{19.88\%}   \\ \bottomrule
\end{tabular}
}
\end{center}
\end{table}

\paragraph{2) Results.}
Figure \ref{fig:module_log} shows the trend of average percentage of retained convolution kernels in a module and the validation accuracy of the \textit{CM} along with training epochs.
In this example, the corresponding \textit{TM} is obtained by training SimCNN on CIFAR-10, and the validation accuracy of the \textit{TM} on the validation dataset is 89.29\%. 
As shown in the left sub-figure in Figure \ref{fig:module_log}, 
modules retain all of the kernels at the beginning of training because masks are initialized with random positive values. GradSplitter also tried to initialize masks using random real and random negative values.
Compared to the initialization with positive values, initialization with random real values could lead to the removal of relevant kernels, causing GradSplitter to converge more slowly. Initializing masks with negative values makes it difficult for GradSplitter to train the masks, as all convolutional layer outputs are zeros at the beginning. Consequently, masks are initialized with positive values by default.
During the training process, the average percentage of retained kernels in a module decreases quickly in the first 40 epochs and then gradually converges.
As shown in the right sub-figure in Figure \ref{fig:module_log}, despite the decreasing number of convolution kernels of modules, the validation accuracy of the \textit{CM} is maintained close to the validation accuracy of the \textit{TM}.
As each \textit{head} in a module has only 11 ($10{+}1$) neurons (detailed in Section \ref{subsec:head}), the optimization of randomly initialized \textit{heads} in \textit{CM} is fast, and the validation accuracy of \textit{CM} is close to that of the \textit{TM} at the first epoch. 

Table \ref{tab:module_results_strong} presents the modularization results of \projectName and CNNSplitter on four strong models, with a comparison between the two approaches.
For instance, as shown in the 3rd row, the trained model SimCNN-CIFAR10 achieves a test accuracy of 89.77\% (2nd column) with 4224 kernels (3rd column). \projectName decomposes SimCNN-CIFAR10 into 10 modules, each retaining an average of 41.22\% of the model's kernels (penultimate column). The \textit{CM}, which is composed of the 10 modules and classifies the same classes as the \textit{TM}, obtains a test accuracy of 88.90\%, with a loss of only 0.87\% compared to SimCNN-CIFAR10 (5th column). In contrast, the modules generated by CNNSplitter retain more kernels, averaging 61.96\% (7th column), and the accuracy of the \textit{CM} is lower at 86.07\%, with a loss of 3.70\% (4th column). Compared to CNNSplitter, \projectName achieves improvements in both accuracy and module size, with an increase of 2.83\% (6th column) and a reduction of 20.74\% (last column), respectively.
As shown in the last row, on average, the accuracy of composed models produced by CNNSplitter and \projectName are 89.77\% and 92.09\%, respectively, with the latter achieving an improvement of 2.32\%. 
Compared with the accuracy of the trained model of 92.66\%, the accuracy losses caused by CNNSplitter and \projectName are 2.89\% and 0.58\%, respectively, indicating that \projectName causes much fewer loss of accuracy.
Regarding the module size, the modules generated by both CNNSplitter and GradSplitter are smaller than the models, with only 56.76\% and 36.88\% of kernels retained, respectively, indicating that the module incurs fewer memory and computation costs than models.

\begin{figure}
    \centering
    \includegraphics[width=9cm]{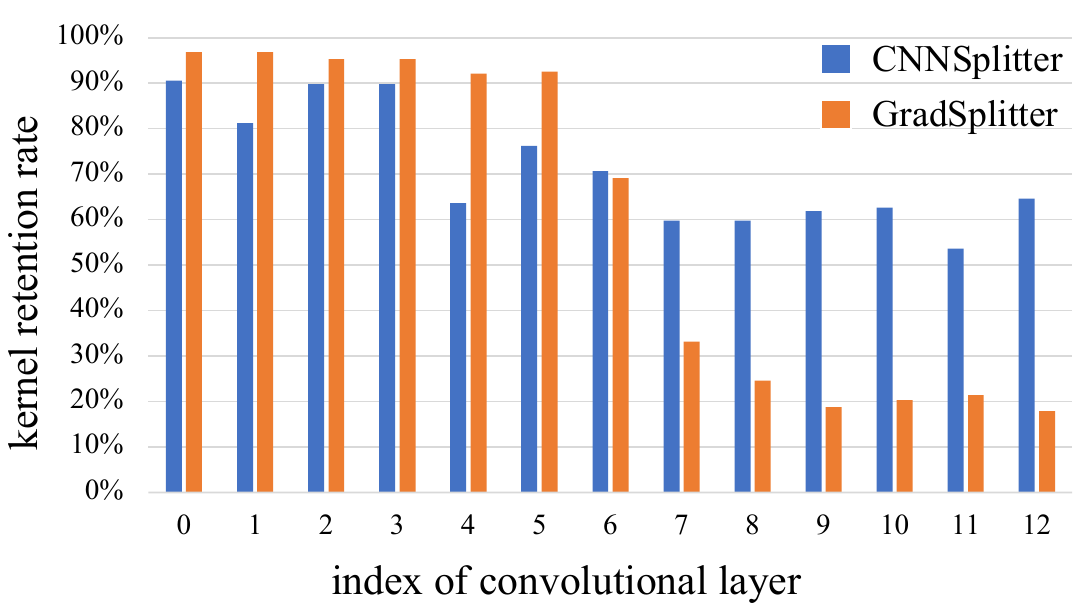}
    \vspace{-6pt}
    \caption{The convolution kernel retention rate for each convolutional layer of SimCNN-CIFAR10.
    }
    \label{fig:distribution}
\end{figure}

We observed that GradSplitter retains fewer convolution kernels than CNNSplitter but incurs less accuracy loss. To explain this outcome, we first analyze the retention of kernels in every convolutional layer and find the difference in kernel retention between modules generated by the two approaches. As shown in Figure \ref{fig:distribution}, GradSplitter retains more kernels in the lower layers (\ie the first 6 convolutional layers) and fewer in the higher layers (\ie the last 7 layers) compared to CNNSplitter. Studies~\cite{transfer3,transfer4} have shown that lower layers (closer to the input layer) learn general features, while higher layers (closer to the output layer) learn specific features of certain classes. Therefore, intuitively, an appropriate distribution of kernel retention for a module should retain more kernels in the lower layers and fewer in the higher layers. GradSplitter performs better than CNNSplitter regarding the distribution of kernel retention, thus retaining fewer kernels while incurring less accuracy loss.

Moreover, the modules generated by GradSplitter have a larger capacity, which is another potential explanation for the outcome.
The capacity of a model can be measured by the number of floating-point operations (FLOPs) required by the model. Larger FLOPs mean that the model has a larger capacity~\cite{gao2021residual,shi2022compressing}. Table~\ref{tab:flops} presents the FLOPs required by the modules and models to classify an image. 
Take the SimCNN-CIFAR10 model as an example (3rd row), on average, a module generated by CNNSplitter requires 164.3 million FLOPs, while a module generated by GradSplitter requires 168.9 million FLOPs. For all four models, the modules generated by GradSplitter require more FLOPs than those generated by CNNSplitter. 
The reason why a module generated by GradSplitter retains fewer kernels but requires more FLOPs is that its lower layers retain more kernels. 
Due to max pooling operations, the inputs of lower layers are larger than those of higher layers, and thus a kernel in the lower layer could incur more FLOPs than a kernel in the higher layer.

\begin{table}[]
\scriptsize
\caption{The FLOPs of the models and decomposed modules.}
\vspace{-6pt}
\label{tab:flops}
\begin{tabular}{cccccc}
\toprule
\multirow{2}{*}{\textbf{Model}} & \multirow{2}{*}{\textbf{\begin{tabular}[c]{@{}c@{}}Model\\ FLOPs (M)\end{tabular}}} & \multicolumn{2}{c}{\textbf{CNNSplitter}}       & \multicolumn{2}{c}{\textbf{GradSplitter}}      \\ \cmidrule(lr){3-4} \cmidrule(lr){5-6} 
                                &                                                                                     & \textbf{Module FLOPs (M)} & \textbf{Reduction} & \textbf{Module FLOPs (M)} & \textbf{Recudtion} \\ \midrule \midrule
SimCNN-CIFAR10                    & \multirow{2}{*}{313.7}                                                              & 164.3                     & 47.60\%             & 168.9                     & 46.16\%             \\
SimCNN-SVHN                     &                                                                                     & 107.8                     & 65.60\%             & 121.2                     & 61.37\%             \\ \midrule
ResCNN-CIFAR10                    & \multirow{2}{*}{431.2}                                                              & 225.4                     & 47.70\%             & 250.2                     & 41.97\%             \\
ResCNN-SVHN                     &                                                                                     & 142.1                     & 67.00\%             & 158.5                     & 63.24\%             \\ \midrule
\textbf{Average}                & \textbf{372.45}                                                                     & \textbf{159.9}            & \textbf{56.98\%}    & \textbf{174.7}            & \textbf{53.18\%}    \\ \bottomrule
\end{tabular}
\end{table}

As shown in Table \ref{tab:flops}, our proposed techniques can significantly reduce the number of FLOPs required by modules, with average reductions of 56.98\% and 53.18\% for CNNSplitter and GradSplitter, respectively.
In contrast, modules produced by uncompressed modularization approaches~\cite{fse2020modularity,nnmodularity2022icse} retain all weights or kernels, resulting in more memory and computation costs.
Since the tools~\cite{moduleToolA, moduleToolB} published by ~\cite{fse2020modularity,nnmodularity2022icse} and our proposed techniques are implemented on Keras and PyTorch, respectively, they cannot directly decompose each other's trained models. 
We attempted to convert PyTorch and Keras trained models to each other; however, the conversion incurs much loss of accuracy (5\% to 10\%) due to differences in the underlying computation of PyTorch and Keras.
Thus, we analyze the open source tools~\cite{moduleToolA, moduleToolB}, including source code files and the experimental data (\eg the trained CNN models and the generated modules). 
The open-source tool keras-flops~\cite{kerasFlops} is used to calculate the FLOPs for the approach described in ~\cite{fse2020modularity}. 
The FLOPs required by a module in ~\cite{fse2020modularity} are the same as those required by the model.
For the project of~\cite{nnmodularity2022icse}, the modules are not encapsulated as Keras model, and there are no ready-to-use, off-the-shelf tools to calculate the FLOPs required by the modules. Therefore, we manually analyze the number of weights of the module and confirm that a module has the same number of weights as the model.
In summary, the experimental results indicate that our compressed modularization approaches outperform the uncompressed modularization approaches~\cite{fse2020modularity,nnmodularity2022icse} in terms of module's size and its computational cost.

\begin{table}[]
\caption{The results of CNNSplitter in different settings.}%
\vspace{-6pt}
\label{tab:cnnsplitter_diff_config}
\scriptsize

\begin{tabular}{ccccr}
\toprule
\multirow{2}{*}{\textbf{Model}} & \multicolumn{2}{c}{\textbf{Settings}} & \multicolumn{2}{c}{\textbf{Composed Model}} \\ \cmidrule(lr){2-3} \cmidrule(lr){4-5}
                                & \textbf{Grouping}       & \textbf{Initialization}     & \textbf{Generation}      & \textbf{Accuracy}        \\ \hline \hline
\multirow{4}{*}{SimCNN-CIFAR10}   & no                      & sensitivity-based                   & 194                      & 0.2754              \\
                                & random                  & sensitivity-based                   & 190                      & 0.3650              \\
                                & importance-based               & random                      & 192                      & 0.3702              \\ \cline{2-5} 
                                & \textbf{importance-based}      & \textbf{sensitivity-based}          & \textbf{123}             & \textbf{0.8607}     \\ \hline
\multirow{4}{*}{SimCNN-SVHN}    & no                      & sensitivity-based                   & 200                      & 0.2430              \\
                                & random                  & sensitivity-based                   & 200                      & 0.2512              \\
                                & importance-based               & random                      & 188                      & 0.9204              \\ \cline{2-5} 
                                & \textbf{importance-based}      & \textbf{sensitivity-based}          & \textbf{79}              & \textbf{0.9385}     \\ \hline
\multirow{4}{*}{ResCNN-CIFAR10}   & no                      & sensitivity-based                   & 83                       & 0.7271              \\
                                & random                  & sensitivity-based                   & 193                      & 0.8420              \\
                                & importance-based               & random                      & 197                      & 0.8432              \\ \cline{2-5} 
                                & \textbf{importance-based}      & \textbf{sensitivity-based}          & \textbf{185}             & \textbf{0.8564}     \\ \hline
\multirow{4}{*}{ResCNN-SVHN}    & no                      & sensitivity-based                   & 140                      & 0.9027              \\
                                & random                  & sensitivity-based                   & 162                      & 0.9249              \\
                                & importance-based              & random                      & 179                      & 0.9332              \\ \cline{2-5} 
                                & \textbf{importance-based}      & \textbf{sensitivity-based}          & \textbf{107}             & \textbf{0.9352}     \\ \bottomrule
\end{tabular}

\end{table}

We also evaluate the effectiveness of importance-based grouping, sensitivity-based initialization, and pruning-based evaluation in CNNSplitter. Table \ref{tab:cnnsplitter_diff_config} shows the results of CNNSplitter under different grouping settings (\textit{no}, \textit{random}, \textit{importance-based}) and initialization settings (\textit{random}, \textit{sensitivity-based}). 
(1) comparing the results under $\langle$importance-based, sensitivity-based$\rangle$ to the results under $\langle$no, sensitivity-based$\rangle$ and $\langle$random, sensitivity-based$\rangle$, we found that modularization without importance-based grouping would cause more accuracy loss and require more generations (e.g., for ResCNN-SVHN) and may even fail due to significant accuracy loss (e.g., for SimCNN-CIFAR and SimCNN-SVHN). 
The time cost for grouping mainly involves collecting the importance of convolution kernels. This process takes about 30 seconds for both SimCNN-CIFAR10 and ResCNN-CIFAR10, and 60 seconds for both SimCNN-SVHN and ResCNN-SVHN. 
(2) comparing the results under $\langle$importance-based, sensitivity-based$\rangle$ to the results under $\langle$importance-based, random$\rangle$, we observed that sensitivity-based initialization could improve search efficiency and reduce accuracy loss. 
Analyzing the sensitivity of convolutional layers takes 169 seconds, 156 seconds, 221 seconds, and 204 seconds for SimCNN-CIFAR10, ResCNN-CIFAR10, SimCNN-SVHN, and ResCNN-SVHN, respectively. 
(3) regarding pruning-based evaluation, in the absence of the pruning strategy, modularization fails due to a timeout (i.e., requires more than several years). In contrast, with the pruning strategy, the time cost per generation for SimCNN-CIFAR, SimCNN-SVHN, ResCNN-CIFAR, and ResCNN-SVHN is 83 seconds, 95 seconds, 80 seconds, and 93 seconds, respectively. 

\begin{figure}
	\centering
	\includegraphics[width=0.7\columnwidth]{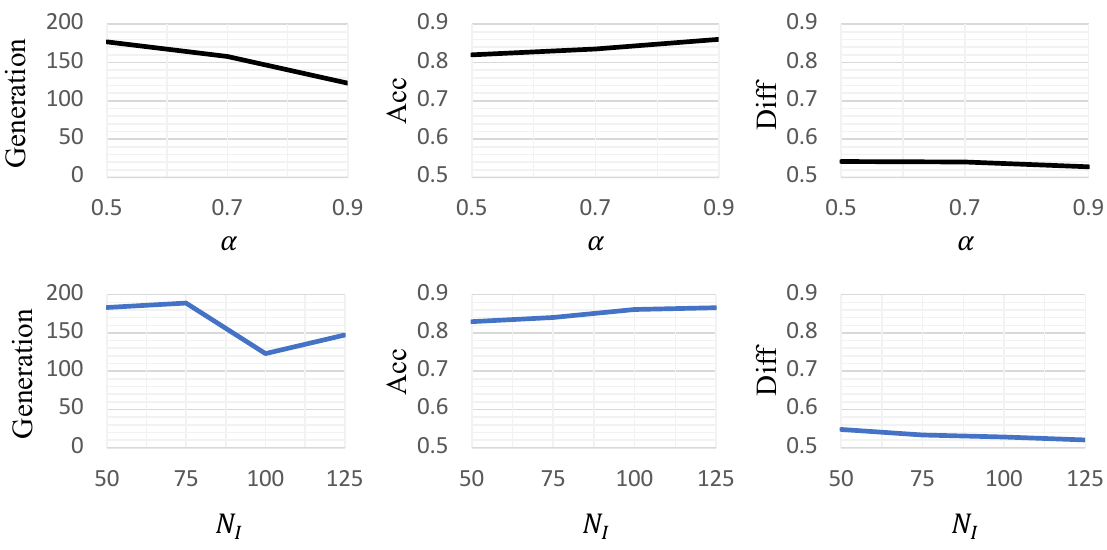}
    \vspace{-6pt}
    \caption{The impact of major parameters in CNNSplitter.}
    \label{fig:impact}
\end{figure}

In addition, we investigate the impact of major parameters on CNNSplitter, including $\alpha$ (the weighting factor between the \textit{Acc} and the \textit{Diff}, described in Sec. \ref{subsec:performance}) and $N_I$ (the number of modules in each generation, described in Sec. \ref{subsec:searchstrategy}). 
Figure \ref{fig:impact} shows the ``Generation'', ``Acc'', and ``Diff'' of the composed model on SimCNN-CIFAR with different $\alpha$ and $N_I$. %
We find that, CNNSplitter performs stably under different parameter settings in terms of \textit{Acc} and \textit{Diff}. The changes in the number of generations show that a proper setting can improve the efficiency of CNNSplitter. 
The results also show that our default settings (\ie $\alpha=0.9$ and $N_I=100$) are appropriate.

\begin{table}[t]
\caption{The impact of $\beta$ in GradSplitter.}%
\centering
\label{tab:alpha}
\footnotesize
\begin{tabular}{cccc}
\toprule
\textbf{$\beta$} & \textbf{Accuracy} & \textbf{Loss of Accuracy} & \textbf{Avg. \#k (\%)} \\ \midrule \midrule
0.01       & 89.19             & 0.58                 & 2297 (54.37)                  \\
0.05       & 88.92             & 0.85                 & 1853 (43.87)                  \\
0.10        & 88.90             & 0.87                 & 1741 (41.22)                  \\
0.50        & 89.12             & 0.65                 & 1947 (46.09)                  \\
1.00          & 89.66             & 0.11                 & 4224 (100.0)  \\ \bottomrule               
\end{tabular}
\end{table}

As for GradSplitter, we investigate the impact of $\beta$ on modularization (see Algorithm \ref{algo:backward}).
Table \ref{tab:alpha} shows the test accuracy and the number of kernels of \textit{CMs} on SimCNN-CIFAR with different $\beta$. 
As \projectName considers the accuracy of \textit{CMs} first when selecting the modules (see Line \ref{algo:overall:select} in Algorithm \ref{algo:overall}), the accuracy of \textit{CMs} with different values of $\beta$ is similar; however, the average number of convolution kernels in a module is different.
As the value of $\beta$ increases from 0.01 to 0.1, the number of kernels decreases, since \projectName prefers to reduce the number of kernels to minimize the weighted loss.
However, the value of $\beta$ should be small, as the excessive value of $\beta$ (\eg $\beta{=}1.0$) could lead to removing kernels dramatically in an epoch, %
resulting in a sharp decrease of accuracy.
The results show that the default $\beta{=}0.1$ is appropriate.

\begin{tcolorbox}
\vspace{-3pt}
Both CNNSplitter and GradSplitter strike a balance between the module’s ability and its size.
In particular, GradSplitter outperforms CNNSplitter, which causes a negligible accuracy loss of 0.58\% and produces smaller modules with only 36.88\% kernels retained.
\vspace{-3pt}
\end{tcolorbox}

\ 

\noindent\textbf{\textit{RQ2: Can the recognition ability of a weak model for a target class be improved by patching?}}

\paragraph{1) Setup.}
\textbf{Design of weak models.}
To answer RQ2, we conduct experiments on three common types of weak CNN models, \ie \textit{overly simple models}, \textit{underfitting models}, and \textit{overfitting models}. 
The modules generated from strong CNN models in RQ1 will be reused to patch these weak CNN models.

An overly simple model has fewer parameters than a strong model.
To obtain the overly simple models, simple SimCNN and ResCNN models are used.
Specifically, a simple SimCNN contains 2 convolutional layers and 1 FC layer, while a simple ResCNN contains 4 convolutional layers, 1 FC layer, and 1 residual connection.

An underfitting model has the same number of parameters as a strong model but is trained with a small number of epochs.
To obtain the underfitting models, the model is trained at the $\frac{n_{best}}{2}$th epoch, which can neither well fit the training dataset nor generalize to the testing dataset. 
The accuracy of the underfitting model is low on both the training dataset and the testing dataset, indicating the occurrence of underfitting. 

An overfitting model is obtained by disabling some well-known Deep Learning ``tricks'', including dropout~\cite{dropout}, weight decay~\cite{weightdecay}, and data augmentation~\cite{dataaugmentation}. These tricks are widely used to prevent overfitting and improve the performance of a DL model. 
The overfitting model can fit the training dataset and the accuracy on the training dataset is close to 100\%; however, its accuracy on the testing dataset is much lower than that on the training dataset, indicating the occurrence of overfitting.
Except for the special design above, the same settings are applied as that of strong models.

\textbf{Training dataset for weak models.}
Considering that the classification of the weak and existing strong models may not be identical in real-world scenarios (which is a reason for patching the weak model rather than directly replacing it with the strong model), weak models are not trained on the same training datasets used by strong models.
For instance, a weak model for the classification of ``cat'' and ``dog'' performs poorly in identifying the class ``cat''. Supposing there is a trained model capable of classifying both ``cat'' and ``fish'' and performs better than the weak model in identifying the class ``cat'', the trained model cannot substitute the weak model but can be used to patch the weak model through modularization.
To construct the training datasets for weak models, a subset of CIFAR-100 and a subset of SVHN are used. The former consists of 9 classes: ``apple'', ``baby'', ``bed'', ``bicycle'', ``bottle'', ``bridge'', ``camel'', ``clock'', and ``rose'', which do not overlap with the CIFAR-10 dataset. Each class in CIFAR-10 is considered as a target class in turn and merged with the subset, resulting in ten 10-class classification datasets, named CIFAR-W. 
The latter consists of 4 fixed classes: ``6'', ``7'', ``8'', and ``9''. Each class of ``0'', ``1'', ``2'', ``3'', and ``4'' is considered as a target class in turn and merged with the subset, resulting in five 5-class classification datasets, named SVHN-W. 
Consequently, fifteen datasets are used to train weak models. The proportion for training, validation, and testing data is 8:1:1.

As a result, 10 CIFAR-W datasets and 5 SVHN-W datasets are constructed to train 3 types of weak models. A total of 90 weak models are obtained, among which, 60 weak models for CIFAR-W (10 datasets with each dataset having 3 weak models for SimCNN and ResCNN, respectively) and 30 weak models for SVHN-W (5 datasets with each dataset having 3 weak models for SimCNN and ResCNN, respectively). 

\textbf{Metrics.}
Given a set of overly simple, underfitting, and overfitting models, the effectiveness of using modules as patches can be validated by quantitatively and qualitatively measuring the improvements of patched models against weak models.
Specifically, the ability of weak models and patched models to recognize a TC can be evaluated in terms of precision and recall. Precision is the fraction of the data belonging to TC among the data predicted to be TC. Recall indicates how much of all data, belonging to TC that should have been found, were found. F1-score is used as a weighted harmonic mean to combine precision and recall. 

\begin{table}[]
\caption{The comparison between GradSplitter and CNNSplitter in patching weak CNN models. The columns ``CS'' and ``GS'' present the performance of patched models based on CNNSplitter and GradSplitter, respectively. The column ``GS-W'' indicates the improvement of patched models based on GradSplitter against weak models.
The column ``GS-CS'' indicates the improvement of GradSplitter against CNNSplitter in patching weak models.  All results in \%.}
\setlength\tabcolsep{2pt}
\label{tab:patching_comparison_all}
\scriptsize
\begin{tabular}{cccccrrcccrrcccrr}
\toprule
\multirow{2}{*}{\textbf{Metric}} & \multirow{2}{*}{\textbf{Model}} & \multicolumn{5}{c}{\textbf{Simple}}                                                                                                                                        & \multicolumn{5}{c}{\textbf{Underfitting}}                                                                                                              & \multicolumn{5}{c}{\textbf{Overfitting}}                                                                                                               \\ \cmidrule(lr){3-7} \cmidrule(lr){8-12} \cmidrule(lr){13-17} 
                                 &                                 & \multicolumn{1}{c}{\textbf{Weak}} & \multicolumn{1}{c}{\textbf{CS}} & \textbf{GS} & \multicolumn{1}{c}{\textbf{GS-W}} & \multicolumn{1}{c}{\textbf{GS-CS}} & \multicolumn{1}{c}{\textbf{Weak}} & \textbf{CS} & \textbf{GS} & \multicolumn{1}{c}{\textbf{GS-W}} & \multicolumn{1}{c}{\textbf{GS-CS}} & \multicolumn{1}{c}{\textbf{Weak}} & \textbf{CS} & \textbf{GS} & \multicolumn{1}{c}{\textbf{GS-W}} & \multicolumn{1}{c}{\textbf{GS-CS}} \\ \midrule \midrule
\multirow{4}{*}{Precision}       & SimCNN-CIFAR                    & 73.30                             & 82.75                                    & 90.05                 & \cellcolor{lightgray}16.75                           & \cellcolor{lightgray}7.30                             & 49.44                             & 70.93                & 76.76                 & \cellcolor{lightgray}27.32                            & \cellcolor{lightgray}5.83                             & 57.34                             & 71.64                & 88.36                 & \cellcolor{lightgray}31.03                            & \cellcolor{lightgray}16.72                            \\
                                 & SimCNN-SVHN                     & 92.32                             & 96.27                                    & 98.34                 & \cellcolor{lightgray}6.02                             & \cellcolor{lightgray}2.07                             & 78.09                             & 91.17                & 98.12                 & \cellcolor{lightgray}20.02                            & \cellcolor{lightgray}6.95                             & 93.76                             & 95.33                & 98.14                 & \cellcolor{lightgray}4.38                             & \cellcolor{lightgray}2.81                             \\
                                 & ResCNN-CIFAR                    & 78.10                             & 89.99                                    & 90.75                 & \cellcolor{lightgray}12.64                            & \cellcolor{lightgray}0.76                             & 45.55                             & 81.66                & 83.22                 & \cellcolor{lightgray}37.67                            & \cellcolor{lightgray}1.56                             & 57.50                             & 76.08                & 75.85                 & \cellcolor{lightgray}18.35                            & -0.23                            \\
                                 & ResCNN-SVHN                     & 89.33                             & 95.86                                    & 98.29                 & \cellcolor{lightgray}8.97                             & \cellcolor{lightgray}2.43                             & 81.39                             & 91.53                & 98.06                 & \cellcolor{lightgray}16.68                            & \cellcolor{lightgray}6.53                             & 92.27                             & 95.60                & 98.05                 & \cellcolor{lightgray}5.78                             & \cellcolor{lightgray}2.45                             \\ \midrule
\multirow{4}{*}{Recall}          & SimCNN-CIFAR                    & 74.50                             & 70.20                                    & 73.60                 & -0.90                            & \cellcolor{lightgray}3.40                             & 36.90                             & 65.00                & 78.10                 & \cellcolor{lightgray}41.20                            & \cellcolor{lightgray}13.10                            & 59.40                             & 57.70                & 61.50                 & \cellcolor{lightgray}2.10                             & \cellcolor{lightgray}3.80                             \\
                                 & SimCNN-SVHN                     & 93.04                             & 91.31                                    & 93.67                 & \cellcolor{lightgray}0.63                             & \cellcolor{lightgray}2.36                             & 77.55                             & 85.96                & 92.16                 & \cellcolor{lightgray}14.61                            & \cellcolor{lightgray}6.20                             & 92.43                             & 91.71                & 92.10                 & -0.33                            & \cellcolor{lightgray}0.39                             \\
                                 & ResCNN-CIFAR                    & 74.30                             & 68.70                                    & 69.90                 & -4.40                            & \cellcolor{lightgray}1.20                             & 39.00                             & 53.60                & 57.30                 & \cellcolor{lightgray}18.30                            & \cellcolor{lightgray}3.70                             & 57.90                             & 55.80                & 55.10                 & -2.80                            & -0.70                            \\
                                 & ResCNN-SVHN                     & 94.68                             & 90.91                                    & 91.59                 & -3.09                            & \cellcolor{lightgray}0.68                             & 84.18                             & 79.14                & 80.00                 & -4.18                            & \cellcolor{lightgray}0.86                             & 93.28                             & 91.79                & 91.61                 & -1.67                            & -0.18                            \\ \midrule
\multirow{4}{*}{F1-score}        & SimCNN-CIFAR                    & 73.76                             & 75.68                                    & 80.57                 & \cellcolor{lightgray}6.81                             & \cellcolor{lightgray}4.89                             & 35.46                             & 64.59                & 77.07                 & \cellcolor{lightgray}41.61                            & \cellcolor{lightgray}12.48                            & 58.14                             & 63.60                & 72.18                 & \cellcolor{lightgray}14.04                            & \cellcolor{lightgray}8.58                             \\
                                 & SimCNN-SVHN                     & 92.67                             & 93.71                                    & 95.94                 & \cellcolor{lightgray}3.27                             & \cellcolor{lightgray}2.23                             & 77.71                             & 88.39                & 95.00                 & \cellcolor{lightgray}17.29                            & \cellcolor{lightgray}6.61                             & 93.08                             & 93.46                & 95.00                 & \cellcolor{lightgray}1.91                             & \cellcolor{lightgray}1.54                             \\
                                 & ResCNN-CIFAR                    & 75.66                             & 77.13                                    & 78.47                 & \cellcolor{lightgray}2.80                             & \cellcolor{lightgray}1.33                             & 34.43                             & 63.05                & 66.95                 & \cellcolor{lightgray}32.52                            & \cellcolor{lightgray}3.90                             & 57.51                             & 64.09                & 63.42                 & \cellcolor{lightgray}5.92                             & -0.67                            \\
                                 & ResCNN-SVHN                     & 91.88                             & 93.30                                    & 94.79                 & \cellcolor{lightgray}2.91                             & \cellcolor{lightgray}1.49                             & 79.63                             & 82.31                & 86.27                 & \cellcolor{lightgray}6.63                             & \cellcolor{lightgray}3.96                             & 92.70                             & 93.61                & 94.65                 & \cellcolor{lightgray}1.95                             & \cellcolor{lightgray}1.04                             \\ \bottomrule
\end{tabular}
\end{table}

\paragraph{2) Results.}
Table \ref{tab:patching_comparison_all} summarizes the precision, recall, and F1-score of weak models and patched models produced by CNNSplitter and \projectName. For instance, in the 3rd row, the 3rd and 5th columns show the average precision of 10 overly simple SimCNN-CIFAR models before and after patching with GradSplitter, respectively. The patched models significantly outperform the weak models (90.05\% vs 73.30\%), representing an improvement of 16.75\% in precision (as shown in the 6th column). Overall, \projectName could improve all types of weak models in terms of precision, with an average improvement of 17.13\%, and generally improves the weak models in terms of recall, with an average improvement of 4.95\%. We observed that the recall values of some patched models decrease (\eg the overly simple SimCNN-CIFAR model), as there is often an inverse relationship between precision and recall~\cite{inverse,fscore}. Nevertheless, the improvement in F1-score indicates that all types of weak models could be improved through patching, with an average improvement of 11.47\%.

We further compare \projectName with CNNSplitter in terms of improvement in recognition of TCs. 
The columns ``GS-CS'' in Table \ref{tab:patching_comparison_all} present the improvements of GradSplitter against CNNSplitter.
Positive improvements are highlighted with a grey background.
Except for the overfitting ResCNN-CIFAR and overfitting ResCNN-SVHN models, the patched models produced by \projectName are superior to those produced by CNNSplitter. Overall, \projectName outperforms CNNSplitter on average by 4.60\%, 2.90\%, and 3.95\% in terms of precision, recall, and F1-score, respectively.
The detailed results in terms of precision, recall, and F1-score are available at the project webpage~\cite{gradsplitter}.

\begin{table}[]
\setlength\tabcolsep{2.8pt}
\caption{
The comparison between GradSplitter and CNNSplitter regarding the accuracy of non-TCs for patched models.
All result in \%.}
\label{tab:patching_comparison_all_nontc}
\scriptsize
\begin{tabular}{cccccccccccccccc}
\toprule
\multirow{2}{*}{\textbf{Model}} & \multicolumn{5}{c}{\textbf{Simple}}                                        & \multicolumn{5}{c}{\textbf{Underfitting}}                                  & \multicolumn{5}{c}{\textbf{Overfitting}}                                   \\ \cmidrule(lr){2-6} \cmidrule(lr){7-11} \cmidrule(lr){12-16} 
                                & \textbf{Weak} & \textbf{CS} & \textbf{GS} & \textbf{GS-W} & \textbf{GS-CS} & \textbf{Weak} & \textbf{CS} & \textbf{GS} & \textbf{GS-W} & \textbf{GS-CS} & \textbf{Weak} & \textbf{CS} & \textbf{GS} & \textbf{GS-W} & \textbf{GS-CS} \\ \midrule \midrule
SimCNN-CIFAR                    & 77.44         & 78.23       & 78.57       & \cellcolor{lightgray}1.12          & \cellcolor{lightgray}0.34           & 42.72         & 43.31       & 43.52       & \cellcolor{lightgray}0.80          & \cellcolor{lightgray}0.21           & 58.59         & 59.47       & 60.07       & \cellcolor{lightgray}1.48          & \cellcolor{lightgray}0.60           \\
SimCNN-SVHN                     & 88.88         & 89.94       & 90.53       & \cellcolor{lightgray}1.65          & \cellcolor{lightgray}0.59           & 73.97         & 75.13       & 76.15       & \cellcolor{lightgray}2.17          & \cellcolor{lightgray}1.02           & 91.14         & 91.52       & 92.19       & \cellcolor{lightgray}1.05          & \cellcolor{lightgray}0.67           \\
ResCNN-CIFAR                    & 81.16         & 81.96       & 82.04       & \cellcolor{lightgray}0.88          & \cellcolor{lightgray}0.08           & 42.80         & 44.53       & 44.61       & \cellcolor{lightgray}1.81          & \cellcolor{lightgray}0.08           & 60.25         & 61.28       & 61.31       & \cellcolor{lightgray}1.06          & \cellcolor{lightgray}0.03           \\
ResCNN-SVHN                     & 88.06         & 90.14       & 90.89       & \cellcolor{lightgray}2.83          & \cellcolor{lightgray}0.75           & 74.99         & 78.82       & 80.52       & \cellcolor{lightgray}5.53          & \cellcolor{lightgray}1.70           & 90.36         & 91.53       & 92.18       & \cellcolor{lightgray}1.82          & \cellcolor{lightgray}0.65           \\ \bottomrule
\end{tabular}
\end{table}

Besides the improvement in recognizing TC, another concern is whether the patch affects the ability to recognize other classes (\ie non-TCs).
To evaluate the patch's effects on non-TCs, the samples belonging to TC are removed, and weak models and patched models are evaluated on the samples belonging to non-TCs.
Finally, the effect of the patch on non-TCs is validated by comparing the accuracy of weak models to patched models. 
The experimental results~\cite{gradsplitter}  are summarized in Table \ref{tab:patching_comparison_all_nontc}.
Overall, when using GradSplitter to patch weak models,  92\% (83/90) of patched models outperform the weak models, and the average accuracy improvement of 90 patched models is 1.85\%. 
The reason for performance improvement is that some samples that belong to non-TCs but were misclassified as TC are correctly classified as non-TCs after patching.
The results indicate that the patching does not impair but rather improves the ability to recognize non-TCs. 
Comparing GradSplitter with CNNSplitter, as shown in the columns ``GS-CS'', 
\projectName performs better than CNNSplitter across all types of weak models, with an average improvement of 0.56\% in non-TCs recognition accuracy.

\begin{tcolorbox}
\vspace{-3pt}
Both CNNSplitter and GradSplitter effectively enhance the recognition performance of weak CNN models on TCs and non-TCs. Notably, GradSplitter outperforms CNNSplitter, which improves weak CNN models in recognizing TCs with an average increase of 17.13\%, 4.95\%, and 11.47\% in precision, recall, and F1-score, respectively.
\vspace{-3pt}
\end{tcolorbox}

\noindent\textbf{\textit{RQ3: Can a composed model, built entirely by combining modules, outperform the best trained model?}}

RQ1 and RQ2 have verified the effectiveness of CNNSplitter and GradSplitter in modularizing CNN models and patching weak models, respectively. Also, the results demonstrate that GradSplitter performs better than CNNSplitter in both modularization and composition. Therefore, in RQ3 to RQ5, we will focus on GradSplitter.
\paragraph{1) Setup.}
\textbf{Dataset construction.}
To investigate whether \projectName can construct better composed models than trained models %
by reusing optimal modules from different models, we conducted experiments on 6 pairs of datasets and CNNs. For each pair, we train 10 CNN models (\textit{TMs}) for 10-class classification and decompose each \textit{TM} into 10 modules.
In practice, the trained models shared by third-party developers could be trained on datasets with different distributions. Therefore, in the experiment, instead of training 10 \textit{TMs} on the initial training set $D$, we draw 10 subsets $\{S_{j}\}_{j=1}^{10}$ from $D$ and train a \textit{TM} on each subset. Each subset $S_{j}=\{S_j^n\}_{n=1}^{10}$ consists of 10 classes of samples, similar to $D=\{D^n\}_{n=1}^{10}$, where $S_j^n$ and $D^n$ indicate the samples in $S_j$ and $D$ belonging to class $n$, respectively.
To ensure that the sampling is reasonable, the sampling is performed according to Dirichlet distribution~\cite{dirichlet}, which is an appropriate choice to make subsets similar to the
real-world data distribution~\cite{lin2020ensemble} and is the hypothesis on which many works are based~\cite{lda,lukins2010bug}. 

Specifically, we first assign the class $n$ a proportion value $p_j^n$ ($0 < p_j^n \leq 1$) that is sampled from the Dirichlet distribution $Dir(\beta)$. 
Then, we draw  $p_j^n \times |D^n|$ samples from $D^n$ randomly to construct $S_j^n$. 
Finally, $S_j$ is constructed once all classes have been sampled. 
Here,  $Dir(\beta)$ denotes the Dirichlet distribution and $\beta$ is a concentration parameter ($\beta > 0$). 
If $\beta$ is set to a smaller value, then the greater the difference between $\{p_j^n\}_{n=1}^{10}$, resulting in a more unbalanced proportion of sample size between the 10 classes. 
For CIFAR-10, we set $\beta=1$, and for SVHN with more data, we set $\beta=0.5$. 
In addition, 
a threshold $t$ is set to ensure that a CNN model has sufficient samples to learn to recognize all classes. 
$p_j^n$ and $S_j^n$ are resampled when $p_j^n \times |D^n|<t$.
We set $t{=}100$ for both CIFAR-10 and SVHN.

Each subset $S_j$ is used to train a \textit{TM} and modularize the \textit{TM}.
Specifically, $S_j$ is randomly divided into two parts in the ratio of $8{:}2$. The 80\% samples are used as \textit{training dataset} to train the \textit{TM} and modularize the \textit{TM}.
The 20\% samples are used as \textit{validation dataset} to evaluate the \textit{TM} and the \textit{CM} during training and modularization.
\label{here}

The initial test dataset is randomly divided into two parts in the ratio of $8{:}2$. 
The 80\% samples are used as the \textit{test dataset} to evaluate \textit{TMs} and \textit{CMs} after training or modularization.
The 20\% samples are used as the \textit{module evaluation dataset} to evaluate and recommend modules (see Section \ref{subsec:module_evaluate}). 

With 6 pairs of datasets and CNNs, we train 10 models for each pair, resulting in 60 CNN models, and then use GradSplitter to decompose these models. The training and modularization settings of the 60 models are the same as that of strong models mentioned in RQ1.

\begin{table*}
\setlength\tabcolsep{2.9pt}
\caption{The modularization results
of \projectName. The columns ``TM'' and ``CM'' present the test accuracy of the \textit{TM} and \textit{CM}. ``\# K'' denotes the average number of kernels in a module.}
\vspace{-6pt}
\label{tab:module_results_all}
\scriptsize
\begin{center}
\begin{tabular}{ccccccccccccccccccc}
\toprule
\multirow{2}{*}{\textbf{Idx}} & \multicolumn{3}{c}{\textbf{SimCNN-CIFAR10}}       & \multicolumn{3}{c}{\textbf{SimCNN-SVHN}}        & \multicolumn{3}{c}{\textbf{ResCNN-CIFAR10}}       & \multicolumn{3}{c}{\textbf{ResCNN-SVHN}}        & \multicolumn{3}{c}{\textbf{InceCNN-CIFAR10}}      & \multicolumn{3}{c}{\textbf{InceCNN-SVHN}}       \\ \cmidrule(lr){2-4} \cmidrule(lr){5-7} \cmidrule(lr){8-10} \cmidrule(lr){11-13} \cmidrule(lr){14-16} \cmidrule(lr){17-19} 
                              & \textbf{TM} & \textbf{CM} & \textbf{\# K} & \textbf{TM} & \textbf{CM} & \textbf{\# K} & \textbf{TM} & \textbf{CM} & \textbf{\# K} & \textbf{TM} & \textbf{CM} & \textbf{\# K} & \textbf{TM} & \textbf{CM} & \textbf{\# K} & \textbf{TM} & \textbf{CM} & \textbf{\# K} \\ \midrule \midrule

0                                        & 79.28                  & \textbf{78.96}         & 1842                           & 86.91                  & \textbf{86.86}         & 1995                           & 80.16                  & \textbf{80.25}         & 1706                           & 85.07                  & \textbf{86.42} & 1679                 & 80.71                  & \textbf{79.81}         & 1275                           & 80.95  & \textbf{81.49}          & 1223                           \\
1                                        & 78.45                  & \textbf{78.29}         & 1933                           & 87.41                  & \textbf{87.68}         & 2002                           & 78.99                  & \textbf{78.11}         & 1903                           & 82.78                  & \textbf{83.50} & 1641                 & 78.26                  & \textbf{77.90}         & 1334                           & 79.06  & \textbf{78.79}          & 1289                           \\
2                                        & 77.80                  & \textbf{77.33}         & 1858                           & 84.45                  & \textbf{83.55}         & 1985                           & 77.53                  & \textbf{77.15}         & 1907                           & 80.96                  & \textbf{79.99} & 2169                 & 79.43                  & \textbf{78.55}         & 1386                           & 80.48  & \textbf{80.05}          & 1452                           \\
3                                        & 80.10                  & \textbf{80.29}         & 2025                           & 82.28                  & \textbf{81.45}         & 2086                           & 81.88                  & \textbf{81.34}         & 2036                           & 80.96                  & \textbf{80.91} & 1828                 & 82.35                  & \textbf{81.81}         & 1543                           & 83.19  & \textbf{82.65}          & 1225                           \\
4                                        & 77.19                  & \textbf{76.61}         & 1913                           & 84.33                  & 81.31                  & 1890                           & 78.09                  & \textbf{77.96}         & 2052                           & 84.54                  & \textbf{84.91} & 1718                 & 81.65                  & \textbf{80.79}         & 1591                           & 81.57  & \textbf{81.14}          & 1388                           \\
5                                        & 79.66                  & 78.49                  & 1938                           & 87.51                  & 85.98                  & 1833                           & 77.93                  & \textbf{77.50}         & 1964                           & 80.45                  & \textbf{79.97} & 2384                 & 79.11                  & 77.90                  & 1513                           & 76.86  & \textbf{77.88}          & 1039                           \\
6                                        & 77.30                  & \textbf{77.09}         & 2043                           & 78.94                  & \textbf{78.68}         & 1930                           & 81.06                  & \textbf{80.95}         & 1977                           & 78.11                  & \textbf{78.51} & 1452                 & 82.70                  & \textbf{82.48}         & 1417                           & 73.03  & \textbf{73.84}          & 1213                           \\
7                                        & 81.01                  & \textbf{80.29}         & 1821                           & 77.75                  & 76.34                  & 2077                           & 80.88                  & \textbf{80.33}         & 1821                           & 74.84                  & \textbf{76.01} & 1842                 & 80.18                  & \textbf{79.58}         & 1383                           & 76.61  & \textbf{76.26}          & 1168                           \\
8                                        & 77.23                  & \textbf{76.76}         & 2035                           & 81.31                  & \textbf{80.55}         & 1773                           & 76.85                  & \textbf{76.62}         & 1946                           & 80.54                  & \textbf{79.85} & 1817                 & 79.66                  & \textbf{78.81}         & 1730                           & 81.19  & \textbf{80.27}          & 1145                           \\
9                                        & 78.20                  & \textbf{77.66}         & 1953                           & 74.82                  & \textbf{73.97}         & 1803                           & 77.00                  & \textbf{76.76}         & 1842                           & 82.75                  & \textbf{82.55} & 1698                 & 83.06                  & \textbf{82.33}         & 1398                           & 69.28  & \textbf{68.37}          & 1203                          

                              \\ \midrule %
\textbf{Avg.}                          & 78.62       & 78.18          & 1936                & 82.57       & 81.64          & 1937                & 79.04       & 78.70          & 1915                & 81.10       & 81.26          & 1823                & 80.71       & 80.00          & 1457                & 78.22       & 78.07          & 1234 \\                  \bottomrule
\end{tabular}
\end{center}
\vspace{-6pt}
\end{table*}

\paragraph{2) Results.}
\textbf{Modularization.}
Table \ref{tab:module_results_all} shows the modularization results of \projectName for six pairs of datasets and CNNs, with 10 \textit{TMs} per pair, for a total of 60 \textit{TMs}. 
Each \textit{TM} is decomposed into 10 modules, and a \textit{CM} is composed of the 10 modules and classifies the same classes as the \textit{TM}.
We evaluate \textit{TMs} and \textit{CMs} on the test dataset and compare the test accuracy of \textit{TMs} and \textit{CMs}.
For each \textit{TM}, Table \ref{tab:module_results_all} shows the test accuracy  of the \textit{TM} and the corresponding \textit{CM}, as well as the average number of kernels in a module (the column ``\# K'').
Among the 60 \textit{CMs}, the 55 \textit{CMs} with less than 1\% accuracy loss are highlighted in bold when compared to \textit{TMs}.

\begin{table}[t]
\caption{The summarized results of modularization. ``Acc.'' denotes the average test accuracy, and ``\# Kernels (\%)'' denotes the average number (and percentage) of kernels in a module.
} 
\vspace{-6pt}
\label{tab:module_results}
\scriptsize
\begin{center}
{
\begin{tabular}{cccccc}
\toprule
\multirow{2}{*}{\textbf{Model}} & \multicolumn{2}{c}{\textbf{Trained Model}} & \multicolumn{3}{c}{\textbf{Composed Model}}   \\ \cmidrule(lr){2-3} \cmidrule(lr){4-6}
                            & \textbf{\# Kernels}       & \textbf{Acc.}       & \textbf{Acc.}     & \textbf{Loss}    &   \textbf{\# Kernels (\%)}                                                                                             \\ \midrule \midrule
SimCNN-CIFAR10                & 4224             & 78.62          & 78.18        & 0.44                & 1936 (46\%)                                                                                    \\
SimCNN-SVHN                 & 4224             & 82.57          & 81.64        & 0.93                & 1937 (46\%)                                                                                    \\ \midrule
ResCNN-CIFAR10                & 4288             & 79.04          & 78.70        & 0.34                & 1915 (45\%)                                                                                    \\
ResCNN-SVHN                 & 4288             & 81.10          & 81.26        & -0.16               & 1823 (43\%)                                                                                    \\ \midrule
InceCNN-CIFAR10               & 3200             & 80.71          & 80.00        & 0.71                & 1457 (46\%)                                                                                    \\
InceCNN-SVHN                & 3200             & 78.22          & 78.07        & 0.15                & 1234 (39\%)                                                                                    \\ \midrule%
\multicolumn{4}{c}{\textbf{Average}}                                              & \textbf{0.40}       & \textbf{1717 (44\%)}                                                                           \\ \bottomrule
\end{tabular}
}
\end{center}
\vspace{-12pt}
\end{table}
Table \ref{tab:module_results} summarizes the results of six pairs of datasets and CNNs, with 10 \textit{TMs} per pair. 
For SimCNN-CIFAR, SimCNN-SVHN, ResCNN-CIFAR, ResCNN-SVHN, InceCNN-CIFAR, and InceCNN-SVHN, the average test accuracy of 10 \textit{TMs} and the average test accuracy
of 10 \textit{CMs} are (78.62\%, 78.18\%), (82.57\%, 81.64\%), (79.04\%, 78.70\%), (81.10\%, 81.26\%), (80.71\%, 80.00\%), and (78.22\%, 78.07\%), respectively. 
As shown in the column ``Loss'', for each pair, the average loss of accuracy for \textit{CMs} compared to \textit{TMs} is less than 1\%.
Overall, the average loss of accuracy of 60 \textit{CMs} is only 0.4\%, which demonstrates that the \textit{CMs} have comparable accuracy to the \textit{TMs} on the 10-class classification tasks. 
The comparable accuracy of the \textit{CMs} suggests that the modules have sufficient ability to recognize the features of the target classes.

We also count the number of convolution kernels in \textit{TMs} and the number of retained convolution kernels in modules. 
The column ``\# Kernels'' of Table \ref{tab:module_results} presents the number of kernels of each \textit{TM}.
And as shown in the last column ``\# Kernels (\%)'' of Table \ref{tab:module_results},
for SimCNN-CIFAR, SimCNN-SVHN, ResCNN-CIFAR, ResCNN-SVHN, InceCNN-CIFAR, and InceCNN-SVHN, a module retains about 40\% convolution kernels of a \textit{TM}.
The average number and percentage of retained convolution kernels in a module for the 60 \textit{CMs} are 1717 and 44\% respectively, indicating that the size of modules is much smaller than that of \textit{TMs}.
The small size of modules leads to a lower prediction overhead than that of \textit{TMs} (as discussed in RQ5). 
Consequently, the modularization results of 60 \textit{TMs} demonstrate that \projectName can strike a balance between the module's recognition ability and the module size. 
The resultant modules will be reused to build better \textit{CMs} than \textit{TMs} (discussed in RQ3) and construct \textit{CMs} for new tasks (as discussed in RQ4).

\textbf{Composition.}
To develop a \textit{CM} that outperforms all \textit{TMs}, modules are first tested on the module evaluation dataset (see Dataset construction in setup). 
Each class in a task is considered in turn as the target class (TC).
For each TC, the corresponding modules that can recognize the TC form a set of candidate modules, 
and \projectName evaluates the candidate modules and recommends the module with the best recognition ability to the developer (detailed in Section \ref{subsec:module_evaluate}).
Candidate modules can come from \textit{TMs} with the same network structure or \textit{TMs} with different network structures. 
We refer to the former as \textit{intra-network reuse} and the latter as \textit{inter-network reuse}.

\begin{table}
\caption{The evaluation results of modules of 10 SimCNN-CIFAR \textit{TMs} in terms of F1-score on module evaluation dataset. The test accuracy of the \textit{TMs} is presented in the last row for comparison.
All results in \%. }
\label{tab:eval_module_intra}
\vspace{-6pt}
\scriptsize
\begin{tabular}{|c|llllllllll|}
\hline
\multicolumn{1}{|l|}{\textbf{\diagbox{TC}{TM}}} & \multicolumn{1}{c}{\textbf{0}} & \multicolumn{1}{c}{\textbf{1}} & \multicolumn{1}{c}{\textbf{2}} & \multicolumn{1}{c}{\textbf{3}} & \multicolumn{1}{c}{\textbf{4}} & \multicolumn{1}{c}{\textbf{5}} & \multicolumn{1}{c}{\textbf{6}} & \multicolumn{1}{c}{\textbf{7}} & \multicolumn{1}{c}{\textbf{8}} & \multicolumn{1}{c|}{\textbf{9}} \\ \hline \hline
\textbf{0}                       & 81.62                 & 76.75                 & 80.60                 & \textbf{\underline{86.27}}        & 77.20                 & 77.73                 & 72.04                 & 84.58                 & 77.88                 & 81.13                 \\
\textbf{1}                       & \textbf{\underline{92.09}}        & 88.67                 & 90.13                 & 90.09                 & 78.82                 & 77.68                 & 82.32                 & 91.04                 & 87.84                 & 85.87                 \\
\textbf{2}                       & 74.03                 & 68.90                 & 74.16                 & 70.74                 & 75.76                 & \textbf{\underline{78.62}}        & 70.53                 & 78.01                 & 50.36                 & 53.42                 \\
\textbf{3}                       & 60.76                 & 59.70                 & 64.15                 & \textbf{\underline{67.18}}        & 64.07                 & 62.84                 & 59.06                 & 66.17                 & 65.09                 & 63.66                 \\
\textbf{4}                       & 70.46                 & \textbf{\underline{81.98}}        & 81.29                 & 69.19                 & 81.73                 & 74.48                 & 62.82                 & 81.15                 & 75.63                 & 80.18                 \\
\textbf{5}                       & 74.33                 & \textbf{\underline{76.78}}        & 60.40                 & 74.93                 & 35.09                 & 67.11                 & 72.88                 & 60.63                 & 71.22                 & 69.18                 \\
\textbf{6}                       & 84.70                 & 74.14                 & 72.78                 & 75.52                 & 84.26                 & \textbf{\underline{86.57}}        & 83.02                 & 83.73                 & 85.32                 & 74.11                 \\
\textbf{7}                       & 76.61                 & 79.17                 & 72.57                 & \textbf{\underline{88.38}}        & 81.92                 & 84.40                 & 80.79                 & 84.40                 & 82.13                 & 84.11                 \\
\textbf{8}                       & \textbf{\underline{89.73}}        & 79.79                 & 88.29                 & 88.74                 & 80.08                 & 88.02                 & 84.28                 & 89.69                 & 80.10                 & 83.06                 \\
\textbf{9}                       & \textbf{\underline{91.61}}        & 83.33                 & 82.94                 & 88.43                 & 76.96                 & 83.26                 & 83.53                 & 81.34                 & 84.26                 & 85.57                 \\ \hline \hline
\textbf{Acc.}                                                                              & 79.28 & 78.45 & 77.80 & 80.10 & 77.19 & 79.66 & 77.30 & \textbf{81.01} & 77.23 & 78.20                           \\ \hline
\end{tabular}
\vspace{-6pt}
\end{table}

\textit{Intra-network reuse}. Table \ref{tab:eval_module_intra} shows the testing results of candidate modules from 10 SimCNN-CIFAR \textit{TMs}, which is used for \textit{intra-network} reuse. 
For each class (TC), there are 10 modules from each of the 10 \textit{TMs}.
Thus, in rows 2 to 11 of Table \ref{tab:eval_module_intra}, each row shows the performance of the modules on the corresponding TC in terms of F1-score,
with the best F1-score highlighted in bold.
The last row shows the accuracy of the 10 \textit{TMs} on the 10-class classification task, with the best accuracy highlighted in bold. 
The testing results show that (1) the \textit{TM} with an index of 7 has the highest accuracy (\ie 81.01\%) on 10-class classification; however, the modules of this \textit{TM} are not optimal in terms of F1-score, and (2) the modules with the best performance for different TCs could come from different \textit{TMs}.

Based on the evaluation results of modules, the modules with the best performance on the module evaluation dataset are reused to build a \textit{CM} for the 10-class classification task. 
We evaluate the \textit{CMs} on the test dataset and compare the test accuracy of \textit{TMs} and \textit{CMs}. 
As shown in the 3rd row of Table \ref{tab:intra_reuse}, the test accuracy of the \textit{CM}, which is composed of the modules highlighted in bold in Table \ref{tab:eval_module_intra}, is 86.26\%. 
The \textit{CM} outperforms the best \textit{TM}, and the improvement in accuracy is 5.25\%. 
Table \ref{tab:intra_reuse} also shows the results for the other five cases of intra-network reuse.
In all six cases, the \textit{CMs} outperform the corresponding best \textit{TMs}. On average, the improvement in terms of accuracy obtained by modularization and composition is 5.22\%.

\textit{Inter-network reuse}. Table \ref{tab:eval_module_inter} shows the testing results of modules that come from 10 SimCNN-CIFAR \textit{TMs} and 10 ResCNN-CIFAR \textit{TMs}. 
For each TC, there are 20 candidate modules. 
Rows 3 to 12 of Table \ref{tab:eval_module_inter} show the performance of the modules in terms of F1-score. 
Similar to intra-network reuse, the evaluation results show that the ResCNN-CIFAR \textit{TM} with index 3 has the highest accuracy (\ie 81.88\%) on 10-class classification; however, not all modules of this \textit{TM} are optimal in terms of F1-score. 
For the 10 TCs, Table \ref{tab:eval_module_inter} highlights the corresponding modules with the highest F1-score.
Among the 10 modules, 5 modules are from the SimCNN-CIFAR \textit{TMs}, and 5 modules are from the ResCNN-CIFAR \textit{TMs}. 
As shown in the 3rd row of Table \ref{tab:inter_reuse}, the test accuracy of the \textit{CM} composed of these 10 modules is 87.24\%. 
Compared to the best \textit{TM}, \ie the ResCNN-CIFAR \textit{TM} with index 3, the improvement in accuracy is 5.36\%. 
Table \ref{tab:inter_reuse} also shows the results of inter-network reuse for the other seven cases. 
In all eight cases, the \textit{CMs} outperform the corresponding best \textit{TMs} with an average improvement of 5.14\% in accuracy.

\begin{table}[t]
\caption{Intra-network reuse. All results in \%.}
\vspace{-6pt}
\label{tab:intra_reuse}
\scriptsize
\begin{center}
{
\begin{tabular}{ccccc}
\toprule
\multirow{2}{*}{\textbf{Dataset}} & \multirow{2}{*}{\textbf{CNN}} & \multicolumn{2}{c}{\textbf{Accuracy}} & \multirow{2}{*}{\textbf{Improvement}} \\ \cmidrule(lr){3-4}
                                  &                                   & \textbf{Best TM}  & \textbf{CM}  &                                       \\ \midrule \midrule
\multirow{3}{*}{CIFAR-10} & SimCNN           & 81.01                       & 86.26                  & 5.25                 \\
                          & ResCNN           & 81.88                       & 85.95                  & 4.07                 \\
                          & InceCNN          & 83.06                       & 86.94                  & 3.88                 \\ \midrule
\multirow{3}{*}{SVHN}     & SimCNN           & 87.51                       & 93.12                  & 5.61                 \\
                          & ResCNN           & 85.07                       & 90.55                  & 5.48                 \\
                          & InceCNN          & 83.19                       & 90.22                  & 7.03                 \\ \midrule %
\multicolumn{4}{c}{\textbf{Average}}                                                                   & \textbf{5.22}       \\ \bottomrule
\end{tabular}
}
\end{center}
\end{table}

\begin{table*}
\setlength\tabcolsep{1.6pt}
\caption{The evaluation results of modules of 10 SimCNN-CIFAR \textit{TMs} and 10 ResCNN-CIFAR \textit{TMs} in terms of F1-score on module evaluation dataset.
The test accuracy of the \textit{TMs} is presented in the last row for comparison.
All results in \%. }
\label{tab:eval_module_inter}
\vspace{-6pt}
\scriptsize
\begin{tabular}{|c|llllllllll|llllllllll|}
\hline
\multirow{2}{*}{\textbf{\diagbox{\textbf{TC}}{\textbf{TM}}}} & \multicolumn{10}{c|}{\textbf{SimCNN-CIFAR}}                                                                                                                                                                                                                                                     & \multicolumn{10}{c|}{\textbf{ResCNN-CIFAR}}                                                                                                                                                                                                                                                              \\ \cline{2-21} 
                                                                 & \multicolumn{1}{c}{\textbf{0}}     & \multicolumn{1}{c}{\textbf{1}}     & \multicolumn{1}{c}{\textbf{2}}     & \multicolumn{1}{c}{\textbf{3}}     & \multicolumn{1}{c}{\textbf{4}}     & \multicolumn{1}{c}{\textbf{5}}     & \multicolumn{1}{c}{\textbf{6}}     & \multicolumn{1}{c}{\textbf{7}}     & \multicolumn{1}{c}{\textbf{8}}     & \multicolumn{1}{c|}{\textbf{9}}     & \multicolumn{1}{c}{\textbf{0}}     & \multicolumn{1}{c}{\textbf{1}}     & \multicolumn{1}{c}{\textbf{2}}     & \multicolumn{1}{c}{\textbf{3}}              & \multicolumn{1}{c}{\textbf{4}}     & \multicolumn{1}{c}{\textbf{5}}     & \multicolumn{1}{c}{\textbf{6}}     & \multicolumn{1}{c}{\textbf{7}}     & \multicolumn{1}{c}{\textbf{8}}     & \multicolumn{1}{c|}{\textbf{9}}     \\ \hline \hline
\textbf{0}        & 81.62                     & 76.75                     & 80.60                     & \textbf{\underline{86.27}}            & 77.20                     & 77.73                     & 72.04                     & 84.58                     & 77.88                     & 81.13                     & 78.37                     & 75.73                     & 81.48                     & 79.59                              & 83.25                     & 75.26                     & 81.98                     & 84.92                     & 72.94                     & 84.01                     \\
\textbf{1}        & 92.09                     & 88.67                     & 90.13                     & 90.09                     & 78.82                     & 77.68                     & 82.32                     & 91.04                     & 87.84                     & 85.87                     & \textbf{\underline{92.57}}            & 89.31                     & 88.83                     & 89.98                              & 81.84                     & 87.82                     & 89.18                     & 86.36                     & 87.07                     & 90.13                     \\
\textbf{2}        & 74.03                     & 68.90                     & 74.16                     & 70.74                     & 75.76                     & \textbf{\underline{78.62}}            & 70.53                     & 78.01                     & 50.36                     & 53.42                     & 72.23                     & 72.11                     & 75.85                     & 69.89                              & 72.80                     & 75.13                     & 74.40                     & 74.01                     & 50.71                     & 53.10                     \\
\textbf{3}        & 60.76                     & 59.70                     & 64.15                     & 67.18                     & 64.07                     & 62.84                     & 59.06                     & 66.17                     & 65.09                     & 63.66                     & 64.77                     & 65.02                     & 62.38                     & \textbf{\underline{72.64}}                     & 64.10                     & 59.83                     & 68.98                     & 66.93                     & 68.83                     & 63.55                     \\
\textbf{4}        & 70.46                     & 81.98                     & 81.29                     & 69.19                     & 81.73                     & 74.48                     & 62.82                     & 81.15                     & 75.63                     & 80.18                     & 66.52                     & 81.40                     & 75.90                     & 82.24                              & 68.10                     & 63.38                     & 61.18                     & \textbf{\underline{85.58}}            & 76.06                     & 80.28                     \\
\textbf{5}        & 74.33                     & \textbf{\underline{76.78}}            & 60.40                     & 74.93                     & 35.09                     & 67.11                     & 72.88                     & 60.63                     & 71.22                     & 69.18                     & 68.24                     & 73.12                     & 61.89                     & 75.28                              & 64.65                     & 72.36                     & 76.14                     & 58.78                     & 70.34                     & 67.65                     \\
\textbf{6}        & 84.70                     & 74.14                     & 72.78                     & 75.52                     & 84.26                     & 86.57                     & 83.02                     & 83.73                     & 85.32                     & 74.11                     & 82.40                     & 78.55                     & 82.17                     & 85.14                              & 86.10                     & 82.11                     & \textbf{\underline{86.82}}            & 85.43                     & 83.26                     & 68.41                     \\
\textbf{7}        & 76.61                     & 79.17                     & 72.57                     & \textbf{\underline{88.38}}            & 81.92                     & 84.40                     & 80.79                     & 84.40                     & 82.13                     & 84.11                     & 79.56                     & 80.00                     & 71.56                     & 86.49                              & 83.87                     & 79.31                     & 82.09                     & 84.68                     & 82.85                     & 83.45                     \\
\textbf{8}        & 89.73                     & 79.79                     & 88.29                     & 88.74                     & 80.08                     & 88.02                     & 84.28                     & 89.69                     & 80.10                     & 83.06                     & \textbf{\underline{89.80}}            & 73.97                     & 89.04                     & 84.29                              & 84.87                     & 86.94                     & 86.03                     & 84.38                     & 79.51                     & 87.91                     \\
\textbf{9}        & \textbf{\underline{91.61}}            & 83.33                     & 82.94                     & 88.43                     & 76.96                     & 83.26                     & 83.53                     & 81.34                     & 84.26                     & 85.57                     & 85.15                     & 80.60                     & 85.20                     & 86.82                              & 83.97                     & 82.87                     & 88.89                     & 83.78                     & 87.23                     & 84.80                     \\ \hline \hline
\multicolumn{1}{|c|}{\textbf{Acc.}}                                   & \multicolumn{1}{c}{79.28} & \multicolumn{1}{c}{78.45} & \multicolumn{1}{c}{77.80} & \multicolumn{1}{c}{80.10} & \multicolumn{1}{c}{77.19} & \multicolumn{1}{c}{79.66} & \multicolumn{1}{c}{77.30} & \multicolumn{1}{c}{81.01} & \multicolumn{1}{c}{77.23} & \multicolumn{1}{c|}{78.20} & \multicolumn{1}{c}{80.16} & \multicolumn{1}{c}{78.99} & \multicolumn{1}{c}{77.53} & \multicolumn{1}{c}{\textbf{81.88}} & \multicolumn{1}{c}{78.09} & \multicolumn{1}{c}{77.93} & \multicolumn{1}{c}{81.06} & \multicolumn{1}{c}{80.88} & \multicolumn{1}{c}{76.85} & \multicolumn{1}{c|}{77.00} \\ \hline
\end{tabular}
\end{table*}
\begin{table}
\caption{Inter-network reuse. All results in \%.}
\vspace{-6pt}
\label{tab:inter_reuse}
\scriptsize
\begin{center}
{
\begin{tabular}{ccccc}
\toprule
\multirow{2}{*}{\textbf{Dataset}} & \multirow{2}{*}{\textbf{CNN}} & \multicolumn{2}{c}{\textbf{Accuracy}} & \multirow{2}{*}{\textbf{Improvement}} \\ \cmidrule(lr){3-4}
                                  &                                   & \textbf{Best TM}  & \textbf{CM}  &                                       \\\midrule \midrule
\multirow{4}{*}{CIFAR-10} & SimCNN-ResCNN         & 81.88                                                                 & 87.24                                                            & 5.36                 \\
                          & SimCNN-InceCNN        & 83.06                                                                 & 87.18                                                            & 4.12                 \\
                          & ResCNN-InceCNN        & 83.06                                                                 & 86.95                                                            & 3.89                 \\
                          & SimCNN-ResCNN-InceCNN & 83.06                                                                 & 87.34                                                            & 4.28                 \\ \midrule
\multirow{4}{*}{SVHN}     & SimCNN-ResCNN         & 87.51                                                                 & 93.35                                                            & 5.84                 \\
                          & SimCNN-InceCNN        & 87.51                                                                 & 93.17                                                            & 5.66                 \\
                          & ResCNN-InceCNN        & 85.07                                                                 & 91.23                                                            & 6.16                 \\
                          & SimCNN-ResCNN-InceCNN & 87.51                                                                 & 93.35                                                            & 5.84                 \\ \midrule %
\multicolumn{4}{c}{\textbf{Average}}                                                                                                                                                                     & \textbf{5.14}                \\ \bottomrule
\end{tabular}
}

\end{center}
\end{table}

Both intra-network reuse and inter-network reuse demonstrate that, using \projectName, a composed CNN model with higher accuracy than all the trained CNN models can be developed through modularization and composition. 
Overall, the average improvement on all 14 cases (6 cases for intra-network reuse and 8 cases for inter-network reuse) is 5.18\%.
Furthermore, comparing intra-network reuse (shown in Table \ref{tab:intra_reuse}) with inter-network reuse (shown in Table \ref{tab:inter_reuse}), we found that the more candidate modules there are, the higher the accuracy of the \textit{CM}.  
For instance, as shown in Table \ref{tab:inter_reuse} (Row 3) and Table \ref{tab:intra_reuse} (Rows 3 and 4), the \textit{CM} composed of the modules from SimCNN-CIFAR \textit{TMs} and ResCNN-CIFAR \textit{TMs} outperforms the \textit{CM} composed of the modules only from SimCNN-CIFAR \textit{TMs} or only from ResCNN-CIFAR \textit{TMs}. 
This suggests that \projectName is promising and can benefit from a large number of shared modules. 

\begin{tcolorbox}
\vspace{-3pt}
Both intra-network and inter-network reuse can develop accurate composed CNN models, and the average accuracy improvement is 5.18\%.
Experimental results demonstrate the feasibility of reusing modules generated by GradSplitter to build more accurate models.
\vspace{-3pt}
\end{tcolorbox}

\ 

\begin{table*}[!h]
\setlength\tabcolsep{1.8pt}
\caption{The 
results of reusing SimCNN modules for developing \textit{CMs} for new tasks. The row ``C'' and column ``S'' indicate the class index of CIFAR10 dataset and SVHN dataset, respectively. All results are test accuracy and are in \%. }
\vspace{-6pt}
\label{tab:reuse_new_simcnn}
\scriptsize
\begin{center}
\begin{tabular}{|c|cc|cc|cc|cc|cc|cc|cc|cc|cc|cc|}
\hline
\textbf{\diagbox{\textbf{S}}{\textbf{C}}} & \multicolumn{2}{c|}{\textbf{0}} & \multicolumn{2}{c|}{\textbf{1}} & \multicolumn{2}{c|}{\textbf{2}} & \multicolumn{2}{c|}{\textbf{3}} & \multicolumn{2}{c|}{\textbf{4}} & \multicolumn{2}{c|}{\textbf{5}} & \multicolumn{2}{c|}{\textbf{6}} & \multicolumn{2}{c|}{\textbf{7}} & \multicolumn{2}{c|}{\textbf{8}} & \multicolumn{2}{c|}{\textbf{9}} \\ \hline \hline
                                                              & \textbf{RTM}    & \textbf{CM}    & \textbf{RTM}    & \textbf{CM}    & \textbf{RTM}    & \textbf{CM}    & \textbf{RTM}    & \textbf{CM}    & \textbf{RTM}    & \textbf{CM}    & \textbf{RTM}    & \textbf{CM}    & \textbf{RTM}    & \textbf{CM}    & \textbf{RTM}    & \textbf{CM}    & \textbf{RTM}    & \textbf{CM}    & \textbf{RTM}    & \textbf{CM}    \\ \cline{2-21}
\textbf{0}                                                    & 99.95          & 99.09          & 100.0          & 98.91          & 99.85          & 96.54          & 99.37          & 92.06          & 99.90          & 99.50          & 99.66          & 96.33          & 99.90          & 99.45          & 99.90          & 99.04          & 99.80          & 99.31          & 99.85          & 99.54          \\
\textbf{1}                                                    & 99.76          & 98.63          & 99.95          & 99.06          & 99.81          & 95.19          & 99.66          & 93.12          & 99.74          & 99.18          & 99.71          & 97.35          & 99.92          & 98.91          & 99.81          & 98.75          & 99.84          & 99.10          & 99.97          & 99.41          \\
\textbf{2}                                                    & 99.74          & 98.46          & 99.84          & 98.77          & 99.74          & 94.65          & 99.59          & 91.94          & 99.87          & 99.01          & 99.62          & 96.83          & 99.81          & 99.49          & 99.81          & 98.70          & 99.68          & 98.30          & 99.94          & 99.18          \\
\textbf{3}                                                    & 99.42          & 96.87          & 99.78          & 97.19          & 99.33          & 94.06          & 98.29          & 86.10          & 99.60          & 98.13          & 99.78          & 90.77          & 99.85          & 98.21          & 99.75          & 97.77          & 99.78          & 97.92          & 99.85          & 99.19          \\
\textbf{4}                                                    & 99.76          & 96.41          & 99.92          & 98.19          & 99.75          & 96.34          & 99.15          & 91.14          & 99.24          & 98.61          & 99.61          & 94.72          & 99.76          & 98.82          & 99.72          & 98.05          & 99.88          & 97.78          & 99.88          & 98.89          \\
\textbf{5}                                                    & 99.80          & 98.31          & 99.92          & 98.82          & 99.67          & 95.22          & 99.56          & 89.00          & 99.71          & 98.97          & 99.64          & 95.46          & 99.88          & 99.00          & 99.71          & 99.05          & 99.79          & 98.74          & 99.84          & 98.89          \\
\textbf{6}                                                    & 99.86          & 98.48          & 99.95          & 98.61          & 99.72          & 96.63          & 99.45          & 92.23          & 99.68          & 99.16          & 99.64          & 95.51          & 100.0          & 99.32          & 99.86          & 98.86          & 99.91          & 98.17          & 100.0          & 99.32          \\
\textbf{7}                                                    & 99.81          & 98.19          & 99.95          & 98.56          & 99.72          & 93.26          & 99.54          & 91.36          & 99.76          & 98.81          & 99.77          & 96.77          & 100.0          & 99.17          & 99.81          & 98.56          & 99.86          & 99.13          & 100.0          & 99.13          \\
\textbf{8}                                                    & 99.85          & 97.56          & 99.95          & 98.40          & 99.75          & 96.63          & 99.07          & 84.18          & 99.71          & 98.31          & 99.76          & 93.14          & 100.0          & 98.67          & 99.80          & 98.46          & 100.0          & 98.77          & 100.0          & 98.92          \\
\textbf{9}                                                    & 99.64          & 98.36          & 99.90          & 98.84          & 99.46          & 94.63          & 99.38          & 87.35          & 99.58          & 98.84          & 99.69          & 93.84          & 99.84          & 99.17          & 99.84          & 98.74          & 99.69          & 98.48          & 99.84          & 98.64          \\ \hline
\end{tabular}
\end{center}
\end{table*}

\begin{table}[]
\caption{The summarized 
 results of reusing modules for developing \textit{CMs} for new tasks. All results are test accuracy and are in \%.}
\label{tab:reuse_new_summary}
\footnotesize
\centering
\begin{tabular}{cccc}
\toprule
\textbf{CNN} & \textbf{RTM} & \textbf{CM} & \textbf{Loss} \\ \midrule \midrule
SimCNN  & 99.75       & 97.10       & 2.65          \\
ResCNN  & 99.75       & 97.22       & 2.53          \\
InceCNN & 99.83       & 97.64       & 2.19          \\ \midrule %
\multicolumn{3}{c}{\textbf{Average}}            & \textbf{2.46} \\ \bottomrule
\end{tabular}
\end{table}
\noindent\textbf{\textit{RQ4: Can a CNN model for a new task be built through modularization and composition while maintaining an acceptable level of accuracy?}}

\paragraph{1) Setup.}
For a new task, if existing models cannot satisfy the functional requirement, a developer can only reuse the model structure and retrain the model from scratch (the retrained model is referred to as \textit{RTM} for short). 
However, through modularization and composition, the modules of existing models can be reused to create a \textit{CM} that satisfies the task without costly retraining. 
Specifically, to develop a \textit{CM} for a new task, each class of the task is treated in turn as TC, and the candidate modules are evaluated on the module evaluation dataset.
Then, for each class, the module that achieves the best classification result is recommended to be reused.
For instance, a new binary classification task is to classify two classes that come from CIFAR-10 and SVHN, respectively. 
The modules from SimCNN-CIFAR \textit{TMs} and SimCNN-SVHN \textit{TMs} (produced in RQ3) can be composed to satisfy the binary classification task. 
Moreover, the test accuracy of a \textit{CM} should be close to that of the \textit{RTM}.
The \textit{RTM} is obtained by retraining SimCNN on the binary classification dataset from scratch with the settings described in RQ1, except that the number of epochs is changed to 30.

\paragraph{2) Results.}
Table \ref{tab:reuse_new_simcnn} shows the test accuracy of \textit{RTMs} and \textit{CMs}. 
A binary classification task is formed by one of CIFAR-10's 10 classes and one of SVHN's 10 classes, for a total of 100 binary tasks.
\textit{CM} refers to the composed model constructed by the best module from the 10 SimCNN-CIFAR models and the best module from the 10 SimCNN-SVHN models. 
\textit{RTM} refers to the binary classification SimCNN trained from scratch. 
The time cost caused by training a binary model from scratch is 150 seconds for SimCNN and ResCNN (calculated by 5 seconds per epoch times 30 epochs), and 180 seconds for InceCNN (calculated by 6 seconds per epoch times 30 epochs).
Table \ref{tab:reuse_new_summary} summarizes the results of reusing SimCNN, ResCNN, and InceCNN modules, showing the average test accuracy of 300 \textit{RTMs} and 300 \textit{CMs}, respectively. 
The results show that the accuracy of \textit{CMs} is slightly lower than that of \textit{RTMs}.
Compared to retraining a model from scratch, reusing modules causes an average accuracy loss of 2.46\%. 

The loss of test accuracy could be due to a slight decrease in module recognition ability caused by modularization (see RQ1).
On the other hand, since each module in a \textit{CM} has not processed the samples belonging to non-TC during training the \textit{TM}, the modules could misclassify non-TC during prediction. For example, the SimCNN-CIFAR modules have not processed the samples from SVHN during training SimCNN-CIFAR models and thus could misclassify.
Overall, compared to the model trained from scratch, the composed CNN models can achieve similar accuracy. 
The experimental results demonstrate that reusing modules to build a new CNN model for a new task is feasible.

\begin{tcolorbox}
\vspace{-3pt}
Compared to the models retrained from scratch, the composed models can achieve similar accuracy (the average loss of accuracy is only 2.46\%).
Experimental results demonstrate that it is feasible to reuse modules to build %
CNN models for new tasks.
\vspace{-3pt}
\end{tcolorbox}

\ 

\begin{table}[]
\caption{Overhead for modularization.}
\label{tab:overhead_modularization}
\scriptsize
\begin{tabular}{cccrr}
\toprule
\multirow{2}{*}{\textbf{Dataset}} & \multirow{2}{*}{\textbf{CNN}} & \multicolumn{2}{c}{\textbf{Time}}                   & \multirow{2}{*}{\textbf{\begin{tabular}[c]{@{}c@{}}GPU\\ Memory\end{tabular}}} \\ \cmidrule(lr){3-4}
                                  &                                 & \textbf{time per epoch} & \textbf{total time} &                                      \\ \midrule \midrule
\multirow{3}{*}{CIFAR-10}         & SimCNN                          & 28 sec                  & 67 min                    & 3.9 GB                               \\
                                  & ResCNN                          & 32 sec                  & 77 min                    & 5.7 GB                               \\
                                  & InceCNN                         & 56 sec                  & 136 min                   & 8.9 GB                               \\ \midrule
\multirow{3}{*}{SVHN}             & SimCNN                          & 33 sec                  & 79 min                    & 3.9 GB                               \\
                                  & ResCNN                          & 36 sec                  & 88 min                    & 5.7 GB                               \\
                                  & InceCNN                         & 57 sec                  & 138 min                   & 8.9 GB                               \\ \midrule %
\multicolumn{2}{c}{\textbf{Average}}                                   & \textbf{40 sec}         & \textbf{98 min}           & \textbf{6.1 GB}                       \\ \bottomrule
\end{tabular}

\vspace{-6pt}
\end{table}

\noindent\textbf{\textit{RQ5: How efficient is \projectName in modularizing CNN models and how efficient is the composed CNN model in prediction?}}

Besides the accuracy of \textit{CMs}, the modularization efficiency of \projectName and the prediction efficiency of \textit{CMs} are also of concern. 
Specifically, decomposing a trained CNN model should not bring too much time and computational overhead to developers. %
Also, the composed CNN models should not incur too much additional overhead than the trained CNN models in the prediction phase.

\textbf{Overhead for Modularization.} Table \ref{tab:overhead_modularization} shows the overhead for decomposing a trained CNN model, including time overhead and computational overhead (\ie the GPU memory usage). 
The time and computational overhead is brought by training masks and heads.
For each pair of dataset and CNN, there are 10 trained CNN models (introduced in RQ3); thus, the values of time overhead and the GPU memory usage presented in Table \ref{tab:overhead_modularization} are the average values of 10 models. 
On average, the time overhead per epoch is 40 seconds, and the total time for decomposing a trained CNN model is 98 minutes. 
The time overhead on SVHN is larger than that on CIFAR-10, as the former has more data than the latter.
For different CNNs, the time overhead for modularization is positively correlated with the number and the size of convolution kernels in the CNN. 
ResCNN has more convolution kernels than SimCNN; thus, the time overhead for ResCNN is larger. 
Since the convolution kernel size of InceCNN is $5{\times}5$, which is larger than that of SimCNN and ResCNN (\ie $3{\times}3$), the time overhead for InceCNN is the largest.
The GPU memory usage is also positively correlated with the number and the size of convolution kernels; thus, the GPU memory usage for ResCNN is larger than SimCNN, and the GPU memory usage for InceCNN is the largest. 
On average, the GPU memory usage is 6.1GB.
The experimental results demonstrate that both time overhead and computational overhead are not expensive. %

Moreover, modularization can be a one-time and offline process performed by the model sharer. With the shared modules, third-party developers could build models without costly training. For instance, modularization on SimCNN-CIFAR10 and SimCNN-SVHN takes approximately 146 minutes (calculated by $67 min + 79 min$). The resulting modules can be used to build 100 binary classification CMs without additional training. In contrast, training 100 binary classification SimCNN models from scratch would take about 250 minutes (calculated by $150s \times 100 \div 60$). Although training a single binary classification model is cost-effective, the total time costs could be significant when dealing with a large number of models. For model sharing platforms with a massive user base, the value generated by avoiding retraining through modularization may be even more substantial.

\textbf{Overhead for Reusing.}
The overhead for reusing refers to the prediction overhead of the composed models (\textit{CMs}), which consists of time overhead and GPU memory usage.
Since the time overhead for predicting a single input is very small, it is measured using all test data. The GPU memory usage comes from two sources: the weights of a \textit{TM}/\textit{CM} and the intermediate results, which can be measured using the open-source tool \textit{torchsummary}~\cite{torchsummary}.
Since each module can be reused as a binary classifier, there are two prediction modes for the \textit{CM}, including parallel prediction and serial prediction.
In parallel prediction, the modules within a \textit{CM} predict simultaneously, while in serial prediction, the modules predict sequentially (one after the other). 
The time overhead and GPU memory usage for the \textit{CM} prediction vary in different modes.

Table \ref{tab:overhead_inference_parallel} shows the time overhead and GPU memory usage for \textit{TMs} and \textit{CMs} in parallel prediction. 
As a module retains only about 44\% convolution kernels (shown in Table \ref{tab:module_results}), the time overhead of a module is less than that of the corresponding \textit{TM}. 
In parallel prediction, a \textit{CM} predicts through executing the modules in parallel; thus, the time overhead of a \textit{CM} is less than that of the corresponding \textit{TM}. 
For all pairs of datasets and CNNs, the \textit{CM} incurs less time overhead than the \textit{TM}. 
On average, 
reusing modules reduces the average prediction time overhead by 28.6\%.
For the GPU memory usage, a \textit{CM} requires more GPU memory than the corresponding \textit{TM}, as the former has more convolution kernels and intermediate results. 
On average, the GPU memory usage for a \textit{TM} and a \textit{CM} is 524.5MB and 4105.1MB, respectively. 
And reusing modules increases the average GPU memory usage by 684.8\%. 

Table \ref{tab:overhead_inference_serial} shows the time overhead and GPU memory usage for \textit{TMs} and \textit{CMs} in serial prediction. 
Though a module incurs less time overhead than the corresponding \textit{TM}, 10 modules predicting in serial incurs more time overhead than the corresponding \textit{TM}. 
On average, %
reusing modules increases the average prediction time overhead by 770.6\%.
For the GPU memory usage, %
reusing modules increases the average GPU memory usage by 55.88\%. 
A \textit{CM} requires more GPU memory than the corresponding \textit{TM}; however, the additional GPU memory usage is much less than that in parallel prediction.
Since modules are executed in serial, only one module produces intermediate results at each moment. 
Consequently, the GPU memory usage for a \textit{CM} consists of the weights of all modules and the intermediate results of only one module; hence the serial prediction takes much less GPU memory usage than the parallel prediction. 

\begin{table}[]
\caption{Prediction overhead in parallel prediction mode. ``Inc.'' denotes the increased overhead of \textit{CMs} over \textit{TMs}.}
\label{tab:overhead_inference_parallel}
\vspace{-6pt}
\scriptsize
\begin{tabular}{ccrrrrrr}
\toprule
\multirow{2}{*}{\textbf{Dataset}} & \multirow{2}{*}{\textbf{CNN}} & \multicolumn{3}{c}{\textbf{Time (s)}}                           & \multicolumn{3}{c}{\textbf{GPU Memory (MB)}}                       \\ \cmidrule(lr){3-5} \cmidrule(lr){6-8}
                                  &                                 & \textbf{TM} & \textbf{CM} & \textbf{Inc. (\%)} & \textbf{TM} & \textbf{CM} & \textbf{Inc. (\%)} \\ \midrule \midrule

\multirow{3}{*}{CIFAR-10}         & SimCNN                            & 4.0             & 2.6             & -35.0                                  & 405.4            & 3303.4          & 714.9                                  \\
                                  & ResCNN                            & 5.4             & 3.5             & -35.2                                  & 441.1            & 3579.0          & 711.4                                  \\
                                  & InceCNN                           & 9.0             & 7.4             & -17.8                                  & 727.0            & 5949.5          & 718.3                                  \\ \midrule
\multirow{3}{*}{SVHN}             & SimCNN                            & 8.8             & 6.2             & -29.5                                  & 405.4            & 3265.8          & 705.7                                  \\
                                  & ResCNN                            & 12.3            & 7.3             & -40.7                                  & 441.1            & 3194.1          & 624.1                                  \\
                                  & InceCNN                           & 21.8            & 18.9            & -13.3                                  & 727.0            & 5338.6          & 634.3                                  \\ \midrule %
                                  
\multicolumn{2}{c}{\textbf{Average}}                                     & \textbf{10.2}   & \textbf{7.7}    & \textbf{-28.6}                         & \textbf{524.5}   & \textbf{4105.1} & \textbf{684.8}                         \\ \bottomrule

\end{tabular}
\end{table}
\begin{table}[]
\caption{Prediction overhead in serial prediction mode. ``Inc.'' denotes the increased overhead of \textit{CMs} over \textit{TMs}.}
\label{tab:overhead_inference_serial}
\vspace{-6pt}
\scriptsize
\begin{tabular}{ccrrrrrr}
\toprule
\multirow{2}{*}{\textbf{Dataset}} & \multirow{2}{*}{\textbf{CNN}} & \multicolumn{3}{c}{\textbf{Time (s)}}                           & \multicolumn{3}{c}{\textbf{GPU Memory (MB)}}                       \\ \cmidrule(lr){3-5} \cmidrule(lr){6-8}
                                  &                                 & \textbf{TM} & \textbf{CM} & \textbf{Inc. (\%)} & \textbf{TM} & \textbf{CM} & \textbf{Inc. (\%)} \\ \midrule \midrule

\multirow{3}{*}{CIFAR-10}         & SimCNN                            & 4.0             & 37.1                                & 827.3                                  & 405.4                                & 1040.7                              & 156.7                                  \\
                                  & ResCNN                            & 5.4             & 43.4                                & 703.3                                  & 441.1                                & 497.6                               & 12.8                                   \\
                                  & InceCNN                           & 9.0             & 79.6                                & 784.4                                  & 727.0                                & 753.8                               & 3.7                                    \\ \midrule
\multirow{3}{*}{SVHN}             & SimCNN                            & 8.8             & 90.4                                & 926.7                                  & 405.4                                & 1046.9                              & 158.3                                  \\
                                  & ResCNN                            & 12.3            & 89.3                                & 625.9                                  & 441.1                                & 455.3                               & 3.2                                    \\
                                  & InceCNN                           & 21.8            & 186.6                               & 756.1                                  & 727.0                                & 731.2                               & 0.6                                    \\ \midrule 
                                  
\multicolumn{2}{c}{\textbf{Average}}                                     & \textbf{10.2}   & \textbf{87.7}                       & \textbf{770.6}                         & \textbf{524.5}                       & \textbf{754.3}                      & \textbf{55.88}                         \\ \bottomrule

\end{tabular}

\end{table}

In serial prediction, \textit{CMs} based on ResCNN and InceCNN increase the GPU memory usage by 12.8\% and 3.7\% on CIFAR-10 and 3.2\% and 0.6\% on SVHN, respectively. 
Compared to \textit{CMs} based on SimCNN, \textit{CMs} based on ResCNN and InceCNN incur much less GPU memory usage, as the ResCNN modules and InceCNN modules retain fewer weights of the corresponding \textit{TM}. 
SimCNN has 3 FC layers, with only one FC layer connecting to the final convolutional layer having its weights reduced. 
ResCNN and InceCNN both have a single FC layer that connects to the final convolutional layer and has its weights reduced.
Compared to SimCNN modules, ResCNN modules and InceCNN modules retain fewer weights of the corresponding \textit{TM}, resulting in less GPU memory usage. 
The experimental results indicate that \projectName is more suitable for CNNs with fewer FC layers. 
Fortunately, the trend in CNN model design is to reduce the number of FC layers~\cite{li2016pruning}. For instance, the FC layers are replaced with average pooling layers in recent work~\cite{resnet,inception,szegedy2016rethinking}.

When comparing parallel and serial prediction, there is a trade-off between time overhead and GPU memory usage.
With sufficient GPU memory, reusing modules in parallel prediction not only achieves better accuracy than the \textit{TM} but also speeds up the prediction.
With limited GPU memory, reusing modules in serial prediction can improve accuracy while incurring only a small increase in GPU memory usage.

\begin{tcolorbox}
\vspace{-3pt}
Modularization only incurs low time and computational overhead.
In prediction, \textit{CMs} incur additional overhead; however, running modules in parallel can balance the time overhead and GPU memory usage. In the case of all modules running in parallel, the time overhead of \textit{CMs} can be significantly lower than that of \textit{TMs}.
\vspace{-3pt}
\end{tcolorbox}

\section{Discussion}
This section discusses the difference between compressed and uncompressed modularization techniques, as well as the threats to the validity of the proposed approaches and experimental results.

\subsection{{Compressed Modularization \textit{vs.} Uncompressed Modularization}}
\label{subsec:discuss}

\begin{figure}[t]
	\includegraphics[width=0.6\columnwidth]{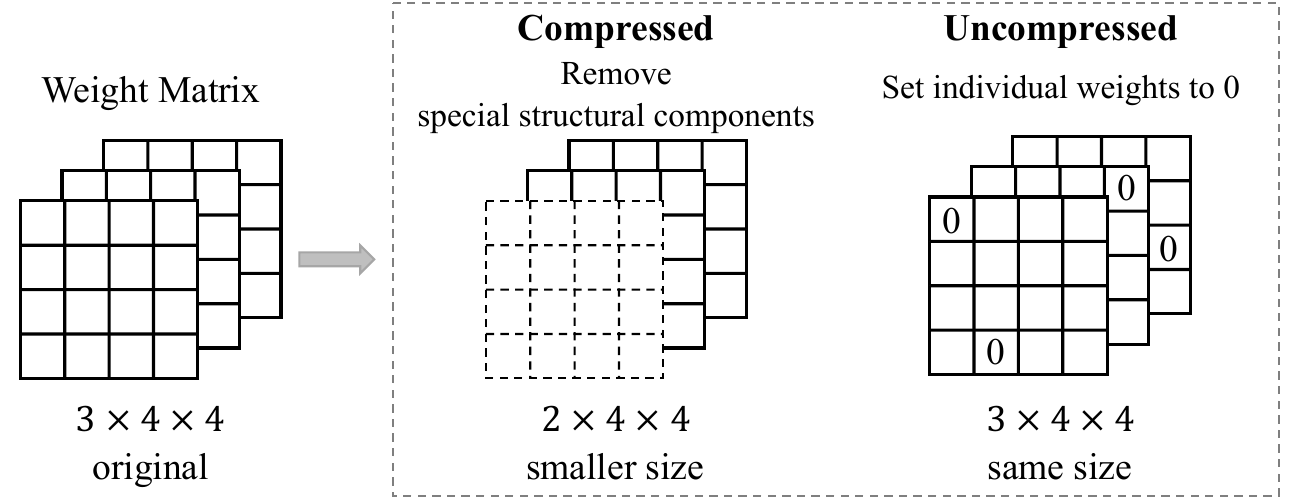}
    \caption{An illustration of the two types of modularization}
    \label{fig:matrix}
    \vspace{-6pt}
\end{figure}

CNNSplitter and \projectName are \textit{compressed modularization} approaches that remove convolution kernels, instead of individual weights, from a trained CNN model (\textit{TM}).
The module generated by compressed modularization has fewer weights than the \textit{TM}.
On the contrary, \textit{uncompressed modularization}~\cite{fse2020modularity, nnmodularity2022icse} removes individual weights or neurons from a \textit{TM} and generates modules with sparse weight matrices (as illustrated in Figure \ref{fig:matrix}).
As a result, a module has the same number of weights as the \textit{TM}, and
the prediction overhead of a \textit{CM} could be significantly higher than that of the \textit{TM} due to much more weights.

To understand the limitations of uncompressed modularization, we analyze the open source project~\cite{moduleToolB} published by Pan \etal~\cite{nnmodularity2022icse}, including source code files and the experimental data (\eg the trained CNN models and the generated modules).
Based on the default settings and the published experimental data, we run the project to evaluate how much additional prediction overhead the \textit{CM} requires compared to the \textit{TM}.
In that project, the test dataset is CIFAR-10, and the \textit{CM} is composed of 10 modules.
Since the \textit{CM} in that project runs each module serially to predict, we only discuss the overhead for the serial prediction mode.

For the \textit{TM}, the prediction time overhead is 2.2s, while the time overhead of the \textit{CM} is 4542.6s. 
The results show that reusing the modules generated by the uncompressed modularization approach~\cite{nnmodularity2022icse} increases the prediction time overhead by 2,064.8\%.
The significant additional time overhead is caused by (1) serial prediction and (2) the additional operations (\ie setting the values of neurons to zero according to the map).
Since each module in the \textit{CM} has the same number of weights as the \textit{TM}, 10 modules predict serially causing about 10 times (1,000\%) overhead than the \textit{TM}. 
Moreover, given an input image, the output (\ie a vector of neuron values) of each convolutional layer cannot be fed directly to the next layer, as part of neuron values in the output need to be set to zeros.

The computational overhead (\ie GPU memory usage) comes from two sources: the weight matrices in the \textit{TM}/\textit{CM} and the intermediate results. 
As a module has the same number of weights as the \textit{TM}, the additional overhead incurred by more weights is about 1,000\%.
On the other hand, there are more additional intermediate results in the \textit{CM} due to the additional operations,
which incurs more overhead.
Since the additional operations are custom operations (\eg iterating the map) that are not integrated into modules as a layer, the overhead caused by intermediate results cannot be computed using existing open-source tools. 
Complex manual computations are required; thus, the overhead of intermediate results is not discussed here.
Overall, in ~\cite{nnmodularity2022icse}, reusing modules increases the prediction time overhead by 2,064.8\% and the computational overhead by about 1,000\%.

Consequently, 
compressed modularization, such as \projectName proposed in this paper, can decompose a trained CNN model into small modules with fewer weights.
The additional prediction overhead in serial prediction mode is significantly less than that of the uncompressed modularization approach.

\subsection{Threats to Validity}
\textbf{External validity:} Threats to external validity relate to the generalizability of our results. 
First, some CNN models have different structures and sizes compared to SimCNN, ResCNN, and InceCNN. 
The results might not be generalizable to these CNN models. 
For instance, CNNSplitter's performance is subject to the size of search space exponentially, leading to a trade-off between the efficiency and quality of modularization. It remains to be validated whether the default settings, such as the number of groups, can achieve a good balance for different models.
Second, the results are not validated on other datasets for different tasks. 
However, SimCNN, ResCNN, and InceCNN are representative CNN models, and their structures are widely used in various tasks. 
Many different CNN models can be seen as variants of SimCNN, ResCNN, and InceCNN. 
In addition, CIFAR-10, CIFAR-100, and SVHN are representative datasets and are widely used for evaluation in related research~\cite{nnmodularity2022icse,feng2020deepgini}. In our future work, more experiments will be conducted on a variety of CNN models and datasets to alleviate this threat.
Finally, a threat relates to the conclusion that a composed model will outperform trained models. In our experiments, the performance of models trained on different sampled distributions could vary (See RQ3); thus, a composed model could outperform trained models by combining their advantage in classifying different classes. However, if models are trained on a similar distribution, the conclusion may not necessarily hold.

\textbf{Internal validity:} %
An internal threat comes from the implementation of trained CNN models. The results of modularization and composition could vary in different training settings, such as hyperparameters and training strategies. In our experiments, the training settings follow the common settings~\cite{vgg, resnet, googlenet}, which are widely used. %

\textbf{Construct validity:} In this study, a threat relates to the suitability of our evaluation metrics. 
The accuracy and the number of retained convolution kernels are used as the metrics to evaluate modularization. 
Moreover, the accuracy, time overhead, and GPU memory usage are used to evaluate the composed model.
These metrics have also been used in other related work~\cite{fse2020modularity,nnmodularity2022icse}.

\section{Related Work}

\subsection{{Modularization of Deep Neural Networks}}

Neural network modularization~\cite{palm,conditioncompute1,ste,moe1,moe2,fse2020modularity,nnmodularity2022icse,mwt,seam,lstmmodular,ye2023mplug} has attracted increasing interest in both the AI and SE communities. We classify existing works into three categories: modularizing \textit{before}, \textit{while}, and \textit{after} training. \textit{Modularizing before training}~\cite{palm,conditioncompute1,ste,moe1,moe2,ye2023mplug} refers to adopting a modular design during the phase of building model architecture. For instance, Mixture-of-Experts (MoE)~\cite{moe1,moe2,moe3} is a typical modular design. The MoE layer consists of multiple experts (similar to modules), each being a fully connected neural network. It also has a gating network that selects a combination of experts to process each input. Through such modular design, these works aim to significantly increase model capacity without a proportional increase in computational overhead.

Different from the aforementioned work, our work belongs to \textit{modularizing after training}~\cite{fse2020modularity,nnmodularity2022icse,seam,lstmmodular}, which focuses on decomposing a trained DNN model into modules, with each module responsible for one subtask of the trained model. 
Existing approaches~\cite{fse2020modularity,nnmodularity2022icse,lstmmodular} design heuristic metrics based on neuron activation to indirectly measure the relevance between weights (or neurons) and sub-tasks. They then remove irrelevant neurons or weights from the model by setting them as zeros to generate modules.
These modularization approaches could have limitations in practice, as the size of the weight matrix of a module is the same as that of the original trained model, resulting in a composed model incurring several times higher prediction overhead than the original model.
In contrast, our work is search-based, which allows for a more direct measurement of the relevance between weights and subtasks according to the impact of weight removal on modules' performance for subtasks. Moreover, our work realizes modularization with reduced network size, helping to decrease the overhead of module reuse.

In addition, Qi et al.~\cite{mwt} recently propose a new paradigm of modularization, namely \textit{modularizing while training} (MwT). MwT introduces the concepts of cohesion and coupling from modular development into the process of training common DNN models from scratch. It designs cohesion and coupling loss functions and guides the training process to follow the modular criteria of high cohesion and low coupling. In contrast, our work belongs to modularizing after training, focusing on decomposing trained models.

\subsection{{Reusing Deep Neural Networks}} 
Our work is related to the work on model reuse, such as direct reuse~\cite{wu2021model, icse21discriminiate} and transfer learning~\cite{transfer_learning, devlin2018bert,transfer3, transfer4}.
Direct model reuse aims to recommend a trained model for developers and enables developers to reuse the model for their new tasks directly.
For instance, Wu \etal~\cite{wu2021model} use the reduced kernel mean embedding (RKME) as a specification for a trained model and then recommend a trained model according to the relatedness of the new task and trained models, which is measured based on the value of the RKME specification.
Transfer learning techniques develop a new model by reusing the entire or a part of a model trained on the other dataset and then fine-tuning the reused model on the new dataset. For instance, BERT~\cite{devlin2018bert} %
can be used to develop new models for various downstream tasks by changing the heads (\ie the output layers) and fine-tuning on the new dataset. 
The techniques mentioned above reuse an entire (or vast majority of) trained model, while our work %
reuses the optimal modules through modularization.

\subsection{DNN Debugging} 
Existing DNN debugging techniques improve DNNs mainly by providing more training data~\cite{ma2018mode}. 
One of the mainstream DNN debugging techniques is the generation technique~\cite{xie2019deephunter, tian2018deeptest, zhang2018deeproad, chen2016infogan}, which generates new training samples that are similar to the provided input data samples. 
For instance, DeepHunter~\cite{xie2019deephunter} and DeepTest~\cite{tian2018deeptest} generate new images by mutating an original image with metamorphic mutations such as pixel value transformation and affine transformation. 
DeepRoad~\cite{zhang2018deeproad} and infoGAN~\cite{chen2016infogan}, both based on generative adversarial networks, train a generator and a discriminator and then use the generator to generate new images.  
Another popular technique is the prioritization technique~\cite{wang2021prioritizing, feng2020deepgini}, which can find the possibly-misclassified data from massive unlabeled data. 
The possibly-misclassified data, rather than total data, are manually labeled first and added into the training dataset to improve a DNN.
Unlike the existing DNN debugging techniques focusing on retraining models with more training data, this work focuses on patching models without retraining.

\subsection{Neural Architecture Search}
Neural architecture search (NAS) techniques~\cite{genetic2017,genetic2019} construct the optimal neural network structure by searching combinations of network layers, layer connections, activation methods, and so on. %
CNNSplitter searches modules from a trained CNN model.
Apart from their differences in objectives, there are some other differences between CNNSplitter and NAS. 
For instance, genetic CNN~\cite{genetic2017} encodes CNN model architectures into bit vectors and applies a genetic algorithm to search. Each bit of a bit vector represents whether or not a connection between two convolutional layers is required. %
While in CNNSplitter, each bit of bit vectors represents whether a kernel group is retained; thus, the approach of genetic CNN cannot be directly applied to modularization. Moreover, CNNSplitter includes three heuristic methods (see Section \ref{sec:approach_cnnsplitter}) to improve the efficiency of search, which are also different from genetic CNN.

\subsection{DNN Pruning}
DNN pruning techniques~\cite{unstructured1, pruning_hansong,topruning,ticket2019iclr, rosenfeld2021predictability} are employed to remove weights that are not important for the whole task, resulting in a smaller model and reducing the resources and time required for inference on the initial task. For instance, magnitude-based pruning~\cite{topruning,ticket2019iclr, rosenfeld2021predictability, li2016pruning,magnitued_cnn_2}, which is one of the mainstream techniques, iteratively prunes the individual weights~\cite{topruning,ticket2019iclr, rosenfeld2021predictability} or convolution kernels~\cite{li2016pruning,magnitued_cnn_2} with the smallest absolute values and trains the retained weights or kernels to recover from pruning-induced accuracy loss. In contrast, our work removes convolution kernels that are irrelevant to the sub-task (\ie classifying one of all classes) to decompose the model into modules, thus facilitating model development and improvement through module reuse.

\section{Conclusion}
In this work, we explore how a trained CNN model can be decomposed into a set of smaller and reusable modules. The resulting modules can be reused to construct completely different CNN models or more accurate CNN models without costly training from scratch.
We propose two compressed modularization approaches including CNNSplitter and \projectName to address the modularity issue of CNN models with a genetic algorithm and gradient-based optimization, respectively. 
We also propose a module evaluation method to guide module reuse, thus providing a new solution for developing accurate CNN models.
We have evaluated CNNSplitter and \projectName with three representative CNN models on three widely-used datasets. 
The experimental results confirm the effectiveness of our approaches. 

In the future, we will explore search-based modularization on more neural networks, such as Transformers at the granularity of attention heads, and further study the modularization and module reuse on large language models such as GPT. 
Additionally, hierarchical modularization, such as decomposing models based on the semantic hierarchy of ImageNet, where a module could be further decomposed into more granular modules, holds potential. This approach could enhance the flexibility of module reuse and the comprehension of model properties.

Our source code and experimental data are available at \textbf{\url{https://github.com/qibinhang/CNNSplitter}} and \textbf{\url{https://github.com/qibinhang/GradSplitter}}.

\section*{Acknowledgement}
This work was supported partly by National Natural Science Foundation of China under Grant Nos.(61972013, 61932007) and partly by Guangxi Collaborative Innovation Center of Multi-source Information Integration and Intelligent Processing.

\bibliographystyle{ACM-Reference-Format}
\bibliography{reference}

\end{document}